\documentclass[11pt,draftcls,onecolumn]{IEEEtran}%
\usepackage{graphicx}
\usepackage{subfigure}
\usepackage{cite}
\usepackage{color}
\usepackage{booktabs}
\usepackage[cmex10]{amsmath}
\usepackage{amsfonts,amssymb,amsthm,bm}
\usepackage{algorithm}
\usepackage{algorithmicx}
\usepackage{algpseudocode}

\begin{document}
\title{Compressed Sensing for Implantable Neural Recordings Using Co-sparse Analysis Model and Weighted $\ell_1$-Optimization}
\author{Biao~Sun, \IEEEmembership{Member, IEEE}, Wenfeng Zhao, \IEEEmembership{Member, IEEE}, and Xinshan Zhu$^*$, \IEEEmembership{Member, IEEE}
\thanks{B. Sun and X. Zhu are with the School of Electrical Engineering and Automation, Tianjin University, Tianjin, 300072, China. Email: \{sunbiao, xszhu\}@tju.edu.cn.}
\thanks{W. Zhao is with the Department of Electrical and Computer Engineering, National University of Singapore, 117583, Singapore. Email: elezhwf@nus.edu.sg.}
\thanks{This work was supported by the National Natural Science Foundation of China under Grants 61271321, 61473207 and 61401303, the Ph.D. Programs Foundation of the Ministry of Education of China under Grant 20120032110068, and Tianjin Key Technology Research and Development Program under Grant 14ZCZDS F00025.}
}

\maketitle

\begin{abstract}
Reliable and energy-efficient wireless data transmission remains a major challenge in resource-constrained wireless neural recording tasks, where data compression is generally adopted to relax the burdens on the wireless data link. Recently, Compressed Sensing (CS) theory has successfully demonstrated its potential in neural recording application. The main limitation of CS, however, is that the neural signals have no good sparse representation with commonly used dictionaries and learning a reliable dictionary is often data dependent and computationally demanding. In this paper, a novel CS approach for implantable neural recording is proposed. The main contributions are: 1) The co-sparse analysis model is adopted to enforce co-sparsity of the neural signals, therefore overcoming the drawbacks of conventional synthesis model and enhancing the reconstruction performance. 2) A multi-fractional-order difference matrix is constructed as the analysis dictionary, thus avoiding the dictionary learning procedure and reducing the need for previously acquired data and computational resources. 3) By exploiting the statistical priors of the analysis coefficients, a weighted analysis $\ell_1$-minimization (WALM) algorithm is proposed to reconstruct the neural signals. Experimental results on Leicester neural signal database reveal that the proposed approach outperforms the state-of-the-art CS-based methods. On the challenging high compression ratio task, the proposed approach still achieves high reconstruction performance and spike classification accuracy.
\end{abstract}


\IEEEpeerreviewmaketitle

\section{Introduction}
Large-scale, multi-channel extracellular neural recording simultaneously from various brain regions are desired to investigate the neural activities from different neuron ensembles, local circuits and brain networks \cite{stevenson2011advances, berenyi2014large}. Such technological capabilities would advance the understanding of brain functions, and moreover, brain-machine interfaces and translational neurotechnologies would become feasible for sophisticated prosthetic devices and disease treatment. In conventional static recording scenario, large amounts of neural data are generated (on the order of tens of Megabytes per second) and tethered cables or wires are commonly adopted for data streaming purposes. However, its applications would be limited owing to tissue infection for subcutaneous, chronic recording tasks as well as in the neuroscience experiments to study awake and free behaving animals models \cite{schwarz2014chronic, yin2014wireless}. Wireless neural recording devices overcome the above-mentioned limitations and would greatly expand the research and application scenarios.

Nevertheless, wireless neural recording devices would compromise among various system-level considerations, including system complexity, power budget and volume miniaturization, and component-level design aspects, such as neural recording amplifiers and analog-to-digital converters (ADCs), neural signal processors, data transceivers and antenna designs. Arguably, the most challenging component in a wireless neural recording device is a reliable, high-throughput and energy-efficient wireless data link, as the wireless link dominates the system channel count, resolution, and energy-efficiency. Although continuous progress is made on data rate and energy efficiency of RF transceivers, it is still prohibitive to adopt such wireless links due to practical limitations of experimental and clinical procedures. Another straightforward approach is to perform on-chip compression before transmission to relax the bandwidth constraints, such as spike detection based approaches \cite{rodriguez2012low, chae2009128, gosselin2009ultra, gosselin2009mixed} and lossy data compression via DWT \cite{oweiss2007scalable}, etc. These approaches significantly reduce the neural data that needed to be transmitted, yet the compression hardware overhead of the on-chip resources and excessive power consumption cannot be neglected.

Recently, the field of Compressed Sensing (CS) \cite{donoho2006compressed, candes2006near} has shown potential in achieving compression and reconstruction performance comparable to the previous approaches but with simpler hardware resources \cite{zhang2014efficient, suo2014energy, zhang2013extension}. The CS approach requires a set of the random measurements of the original signals, and avoids the need for dedicated DSPs and leaves most of the computational burden to off-chip processing. Its main challenge, however, is that the spike segments are not sparse on common dictionaries such as Discrete Cosine Transform (DCT) basis and Discrete Fourier Transform (DFT) basis. Reconstructing the spikes using these dictionaries will severely degrade the performance. Therefore, a careful design of the sparsifying dictionary is needed to guarantee the compression performance \cite{bulach2012evaluation}. To alleviate this issue, various dictionary-learning based algorithms are proposed for neural data compression \cite{zhang2014efficient, suo2014energy}. Zhang \emph{et al.} \cite{zhang2014efficient} proposed learning dictionaries using K-SVD and developed a signal-dependent CS approach to compress the data. Suo \emph{et al.} \cite{suo2014energy} proposed to use the recorded neural data directly as the sparsity dictionary. However, these algorithms are computational demanding and highly signal-dependent. This indicates that iterative training processes are required during practical neural recording applications, which is unfavorable in most experimental settings.

This paper proposes a novel CS framework for implantable neural recordings that is capable of recovering neural spikes with high compression ratio but avoiding the sparsity dictionary learning procedure. The main contributions of the work are as follows.

i) Instead of using conventional synthesis model in CS, the analysis model is adopted to enforce co-sparsity of neural signals, overcoming drawbacks of conventional model and enhancing the reconstruction performance. To our best knowledge, this is the first time that neural signal reconstruction problem is solved by using the analysis model of CS.

ii) Based on the piecewise smooth structures in neural signals, a multiple-fractional-order-difference matrix is constructed as the analysis dictionary. It not only has high co-sparsity with neural signals but also avoids the dictionary learning procedure, saving both computational resources and data storage space.

iii) The statistical priors of the analysis coefficients among difference orders are deduced. The associated reconstruction algorithm, dubbed weighted analysis $\ell_1$-minimization (WALM), is proposed to improve the reconstruction performance by embedding the multiple orders knowledge within penalty weights.

The remainder of this paper is organized as follows. Section II describes the CS-based implantable neural recording system architecture, the relevant background of synthesis model and co-sparse analysis model. Section III introduces the proposed construction method of the multiple-fractional-order-difference dictionary. Section IV covers the weighted analysis $\ell_1$-minimization algorithm for neural spike reconstruction. In Section V, the experimental results are presented and compared to state-of-the-art CS-based reconstruction methods. Section VI concludes the paper.

Throughout the paper, boldface capital letters (e.g., $\bm{A}$) denote matrices, boldface lowercase letters (e.g., $\bm{x}$) denote vectors, not bold letters (e.g., $c$) denote scalars, and boldface calligraphic letters (e.g., $\bm{\mathcal{I}}$) specify number sets. For a vector $\bm{x}$, we use $x_i$ to denote its $i$th entry, and we use $\|\bm{x}\|_2$, $\|\bm{x}\|_1$, and $\|\bm{x}\|_0$ to indicate its $\ell_2$, $\ell_1$, and $\ell_0$ norms, respectively. For a matrix $\bm{A}$, we use $\bm{A}_i$ to denote its $i$th row or column depending on the situation it is used. For a set $\bm{\mathcal{I}}$, we use $|\bm{\mathcal{I}}|$ to indicate its cardinality. For a random variable $a$, its probability distribution function (\emph{pdf}) is denoted by $p(a)$, and its standard deviation is indicated by $\sigma_a$.

\begin{figure}[tb]
	\centering
	\includegraphics[width=.3\textwidth]{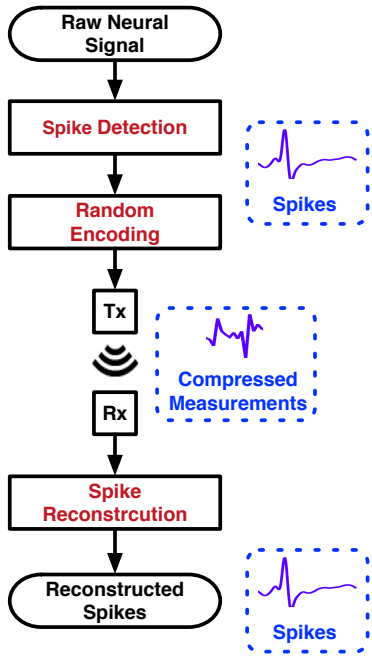}
	\caption{Diagram of the compressed sensing system for implant neural recording.}
	\label{fig:flowchart}
\end{figure}

\section{System Overview and Sparse Models}

\subsection{System Overview}
CS-based wireless neural recording system architecture is briefly depicted in Fig. \ref{fig:flowchart}. The recorded raw neural data are first conditioned into appropriate signal amplitude and bandwidth through amplifiers, filters and digitized via Nyquist-rate ADCs. Second, the neural spike events are detected through threshold crossing techniques and aligned temporally (``Spike Detection'' in the figure).  The aligned segments containing the spikes are then compressed via randomized encoding circuit (``Random Encoding'' in the figure) based on the compressive sensing theory, and the compressed data are transmitted via wireless transmitters (e.g., Bluetooth, Zig-Bee, Wi-Fi). On the receiver side, the random measurements of aligned spikes are reconstructed through some specific algorithms (``Spike Reconstruction'' in the figure) at workstations or fusion centers.

\subsection{Compressed Sensing and Synthesis Model}
Compressed sensing is an emerging low-rate sampling scheme for the signals that are known to be sparse or compressible in certain basis. Assume a signal $\bm{x}\in\mathbb{R}^n$ is measured by a simple matrix-vector multiplication,
\begin{equation}
\label{eq:CS}
\bm{y} = \bm{\Phi x} + \bm{e},
\end{equation}
where $\bm{\Phi}\in\mathbb{R}^{m\times n}$ is called the sensing matrix, $\bm{y}\in\mathbb{R}^m$ is the compressed measurement vector, $\bm{e}\in\mathbb{R}^m$ denotes the measurement noise. Usually Eq. (\ref{eq:CS}) is undetermined, i.e., $m < n$, and the ratio $m/n$ is called the compression ratio of CS. In this undetermined system, the signal $\bm{x}$ cannot be uniquely retrieved from sensing matrix $\bm{\Phi}$ and measurements $\bm{y}$. However, if the $\bm{x}$ can be described using a synthesis model \cite{bruckstein2009sparse}, i.e,
\begin{equation}
\label{eq:basis}
\bm{x} = \bm{\Psi}\bm{\theta},
\end{equation}
where $\bm{\Psi}\in\mathbb{R}^{n\times n}$ is a pre-defined dictionary, and the signal's representation $\bm{\theta}\in\mathbb{R}^n$ is assumed to be $k$-sparse, i.e.,
\begin{equation}
\|\bm{\theta}\|_0 \triangleq |\text{supp}(\bm{\theta})|=k \ll n,
\end{equation}
or is well-approximated by a $k$-sparse vector. The name ``synthesis'' comes from the relation (\ref{eq:basis}), with the obvious interpretation that the model describes a way to synthesize a signal. Therefore, based on Eq. (\ref{eq:CS}) and (\ref{eq:basis}), the compressed measurements $\bm{y}$ can be represented as
\begin{equation}
\bm{y} = \bm{\Phi\Psi\theta} + \bm{n} = \bm{A\theta} + \bm{e},
\end{equation}
where $\bm{A} = \bm{\Phi\Psi}$. Due to the sparse prior knowledge of $\bm{\theta}$, it is possible to estimate $\bm{\theta}$ via the $\ell_0$ minimization formulation as
\begin{equation}
\label{eq:l0}
\bm{\hat{\theta}} = \text{min}\|\bm{\theta}\|_0, \quad \text{s.t.} \quad \|\bm{y}-\bm{A\theta}\|_2^2<\epsilon,
\end{equation}
where $\epsilon$ is the tolerance of noise or modeling errors. Calculating the solution is very
hard because Eq. (\ref{eq:l0}) is an NP-hard problem \cite{candes2006near}. Generally, one seeks the solution of a relaxed convex optimization problem \cite{becker2011nesta}, in which $\|\bm{\theta}\|_0$ is replaced with $\|\bm{\theta}\|_1$ as
\begin{equation}
\label{eq:l1}
\bm{\hat{\theta}} = \text{min}\|\bm{\theta}\|_1, \quad \text{s.t.} \quad \|\bm{y}-\bm{A\theta}\|_2^2<\epsilon.
\end{equation}
Under the condition of Restricted Isometry Property (RIP) \cite{candes2008restricted}, minimizing $\ell_1$ has been theoretically proven to be equivalent to minimizing $\ell_0$. Moreover, $\ell_1$ minimization is convex and can be solved within polynomial time. After estimating the sparse coefficient $\bm{\theta}$, the original signal $\bm{x}$ can be recovered directly by Eq. (\ref{eq:basis}).

\subsection{Co-sparse Analysis Model}
\label{subsection:CAM}
While the synthesis model has been extensively studied, its ``twin model'' that takes an analysis point of view has been left aside almost untouched \cite{elad2007analysis}. The alternative assumes that for a signal of interest, the \emph{analysis coefficients} vector
\begin{equation}
\label{eq:analysismodel}
\bm{z} = \bm{\Omega}\bm{x}
\end{equation}
is expected to be sparse, where $\bm{\Omega}\in\mathbb{R}^{l\times n}$ is a possibly redundant analysis dictionary ($l\geq n$), and the ratio $\rho=l/n$ is called the redundant ratio. The co-sparsity of a signal $\bm{x}$ with respect to $\bm{\Omega}$ is defined as the number of zeros in the vector $\bm{z}$, i.e.,
\begin{equation}
k_\text{co} = l-\|\bm{z}\|_0,
\end{equation}
and the index set of the zero entries of $\bm{z}$ is called the co-support of $\bm{x}$. It is worth noting that for a square and invertible dictionary, the synthesis and the analysis models are the same with $\bm{\Psi} = \bm{\Omega}^{-1}$ \cite{elad2007analysis}. While the analysis model may seem similar to the synthesis counterpart one, it is in-fact very different when dealing with a redundant dictionary $p>n$ \cite{nam2013cosparse}. The traditional synthesis model puts an emphasis on the non-zeros of the sparse vector $\bm{\theta}$, but the co-sparse analysis model draws its strength from the zeros of the analysis vector $\bm{z}$. The optimization problem for co-sparse signal recovery can be formulated as
\begin{equation}
\label{eq:analysisl0}
\bm{\hat{x}} = \text{min}\|\bm{\Omega}\bm{x}\|_0, \quad \text{s.t.} \quad \|\bm{y}-\bm{\Phi x}\|_2^2<\epsilon.
\end{equation}
Here we call (\ref{eq:analysisl0}) the analysis $\ell_0$-minimization. There is also a classical way to relax the nonconvex $\ell_0$ norm into convex $\ell_1$ norm, i.e.,
\begin{equation}
\label{eq:analysisl1}
\bm{\hat{x}} = \text{min}\|\bm{\Omega}\bm{x}\|_1, \quad \text{s.t.} \quad \|\bm{y}-\bm{\Phi x}\|_2^2<\epsilon.
\end{equation}
We call (\ref{eq:analysisl1}) the analysis $\ell_1$-minimization (AL1). Several sufficient conditions theoretically guarantee the successful recovery of the original signal from the compressed measurement using (\ref{eq:analysisl1}), such as the restricted isometry property adapted to the dictionary (D-RIP), restricted orthogonal projection property (ROPP), etc \cite{nam2013cosparse, candes2011compressed, peleg2013performance}.

\section{Analysis Dictionary Construction}
In this section, we focus on the construction of the analysis dictionary $\bm{\Omega}$ so that the analysis coefficients $\bm{\Omega}\bm{x}$ are sparse. It is worth noting that when dealing with a square (and invertible) matrix $\bm{\Omega}$, the analysis model is completely equivalent to the synthesis one, and in such a case, the synthesis-dictionary construction methods can be used to build $\bm{\Omega}$. In this work, we concentrate on the redundant case $l>n$, where the two models depart, and where the analysis model becomes more powerful.

\subsection{Multiple-integer-order-difference Matrix}
Prior works show that many types of signals \cite{liu2015compressed, chambolle2004algorithm}, e.g., EEG and ECG signals, often reveals approximately piecewise smooth structure \cite{keogh2001dimensionality, jiang2009new, zhou2012new, yan2013approach}. This structure exhibits gradient sparsity, i.e., signals will become sparse when differenced with some specific orders. Moreover, investigations of the statistical properties using the available implantable neural signals show that neural spikes are also approximately piecewise smooth, implying that neural spikes fit the co-sparse signal model (\ref{eq:analysismodel}) well with the integer-order-difference (IOD) sequence \cite{et1995some} as analysis dictionary. For an $n$-length signal $\bm{x}$, the IOD sequence of $\bm{x}$ is defined as
\begin{equation}
\Delta^r(\bm{x}) = \sum_{k=0}^{r}(-1)^k\binom{r}{k}x_{i+k},\ i = 1,\dots,n,
\label{IODS}
\end{equation}
where $\Delta^r(\cdot)$ denotes IOD operator and $r\in\mathbb{N}$ is difference order. The IOD sequence can be reformulated into matrix form $\bm{D}^{r}$ as proposed in \cite{et1995some}. For simplicity, we do not distinguish between the two forms in the rest of the paper, and we use matrix form to build the analysis dictionary.

\begin{figure}[tb]
	\centering
	\includegraphics[width=.5\textwidth]{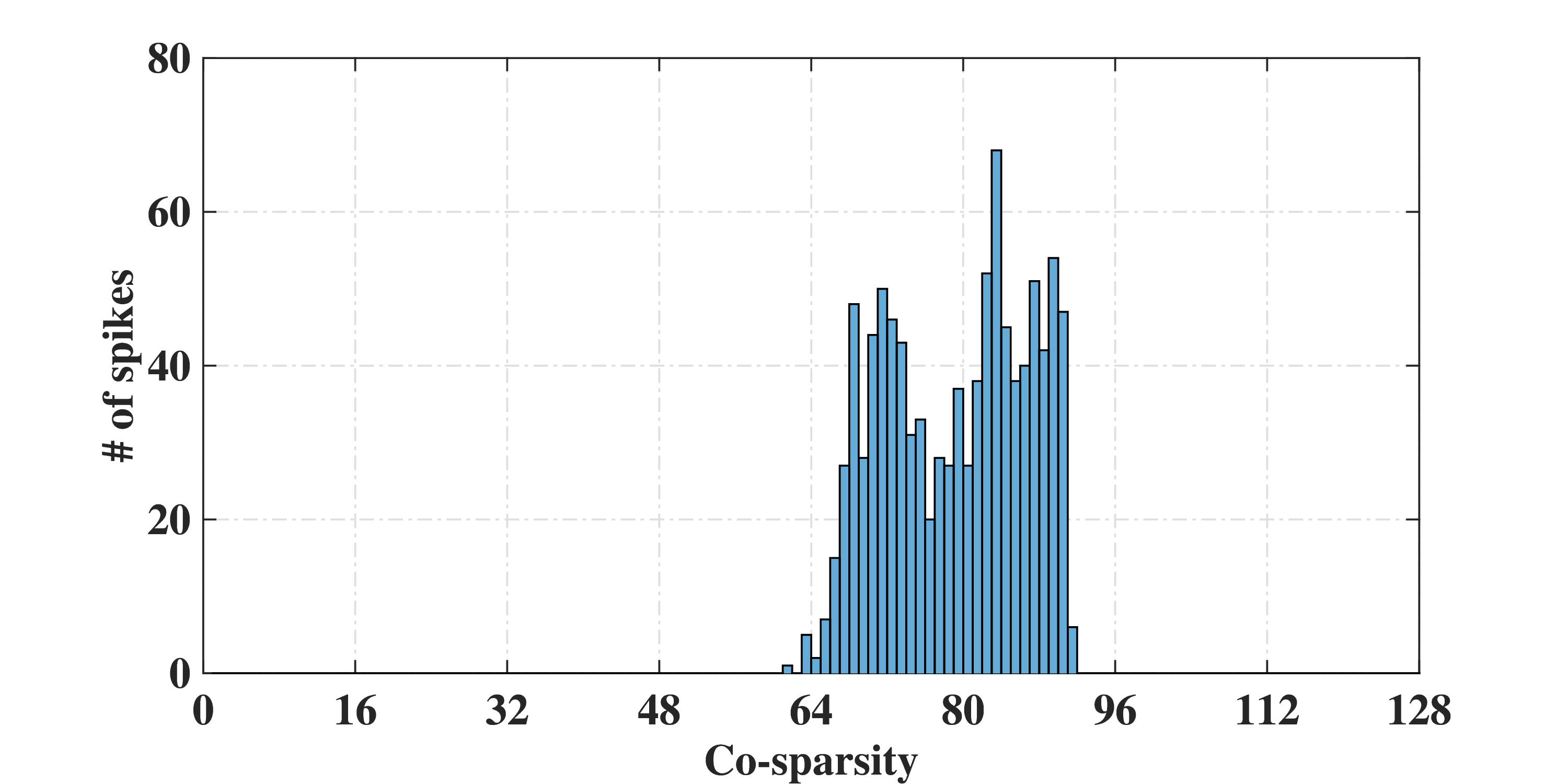}
	\caption{A histogram of the effective co-sparsities of the 1000 spikes using 2nd order IOD matrix as the analysis dictionary. The spikes are randomly chosen from Leicester Easy2 dataset. The length of each spike is $n = 128$.}
	\label{fig:histogram}
\end{figure}

Fig. \ref{fig:histogram} presents a histogram of the co-sparsities of 1000 spikes\footnote{The spikes are randomly chosen from Leicester Easy2 dataset \cite{quiroga2004unsupervised}.} using 2nd order IOD matrix as the analysis dictionary. As can be seen, the co-sparsities are all strictly high. Furthermore, to construct a redundant analysis dictionary, we seek to promote co-sparsity over multiple orders difference sequence rather than a single one, and propose using a multiple-integer-order-difference (MIOD) matrix composed of a concatenation of $q$ IOD matrices, i.e.,
\begin{equation}
\bm{\Omega}_\text{MIOD} \triangleq \frac{1}{\sqrt{q}}[\bm{D}^{r_0},\bm{D}^{r_1},\dots,\bm{D}^{r_{q-1}}]^T,
\label{eq:RAD}
\end{equation}
where $\bm{D}^{r_i}$ denotes the IOD matrix of order $r_i$. The corresponding order set is
\begin{equation}
\bm{\mathcal{R}} = \left\{r_0,r_1,\dots,r_{q-1}\right\}\in\mathbb{N}^{q\times 1}.
\end{equation}
Given the MIOD matrix, we propose the following prior, proportional to the multiple-order sparsity as
\begin{equation}
\|\bm{\Omega}_\text{MIOD}\bm{x}\|_0 = \sum_{r_i\in\bm{\mathcal{R}}}\|\bm{D}^{r_i}\bm{x}\|_0.
\end{equation}
It is worth noting that in this setting the analysis coefficients of each order contain all the signal information, which cannot be formulated in a synthesis model.

\subsection{Multiple-fractional-order-difference Matrix}
\begin{figure}[tb]
	\centering
	\includegraphics[width=.5\textwidth]{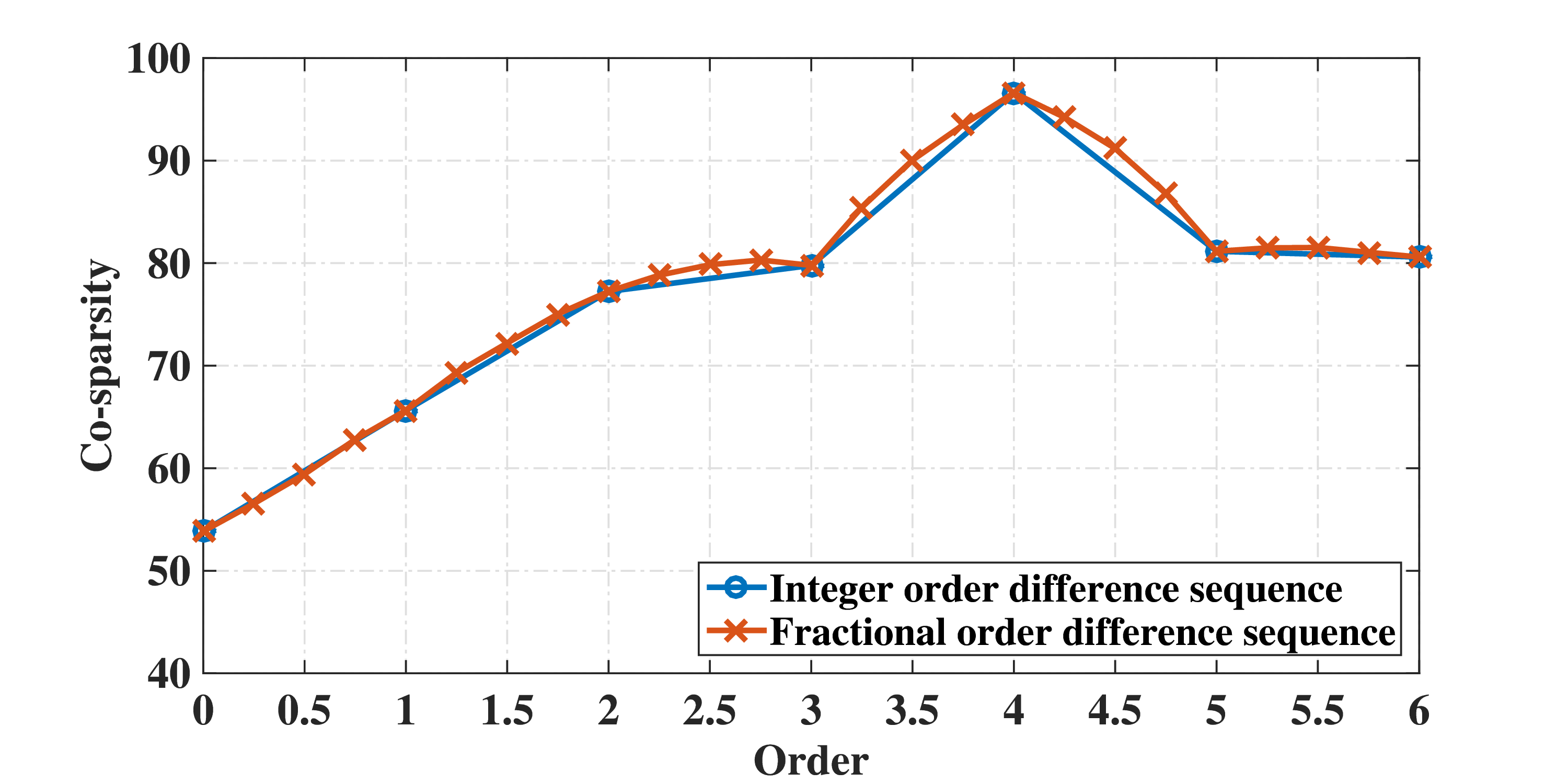}
	\caption{The average co-sparsities of the 1000 spikes using IOD matrix (blue-circle curve) and FOD matrix (red-times curve) as the analysis dictionary, respectively. The spikes are randomly chosen from Leicester Easy2 dataset. The length of each spike is $n = 128$.}
	\label{fig:cosparse}
\end{figure}

Although the MIOD matrix promotes higher co-sparsity than the single-order one, how to choose the order set appropriately remains a problem. The blue-circle curve in Fig. \ref{fig:cosparse} shows the co-sparsity of 1000 spikes using IOD matrix as the analysis dictionary, from which we notice that the 4-th order analysis coefficients have maximum co-sparsity. If adding difference matrix of other orders to $\bm{\Omega}$, however, the co-sparsity of the analysis coefficients will not be optimal, and the signal reconstruction performance will severely degrade.

To solve this problem, we build the multiple-fractional-orders-difference (MFOD) matrix using the fractional-order-difference (FOD) sequence, which is a generalization of IOD sequence to fractional order. The FOD sequence is defined as
\begin{equation}
\Delta^f(\bm{x}) = \sum_{k=0}^{\infty}(-1)^k\frac{\Gamma(f+1)}{k!\Gamma(f-k+1)}x_{i+k},\ i = 1,\dots,n,
\label{eq:FODS}
\end{equation}
where $\Gamma(\cdot)$ denotes the gamma function. It is easy to verify that the summation in (\ref{eq:FODS}) is convergent. If $f$ is a nonnegative integer, then the infinite sum defined in (\ref{eq:FODS}) reduces to a finite sum, i.e.,
\begin{equation}
\Delta^f(\bm{x}) = \sum_{k=0}^{f}(-1)^k\frac{\Gamma(f+1)}{k!\Gamma(f-k+1)}x_{i+k},\ i = 1,\dots,n,
\end{equation}
and this operator generalizes the one defined in (\ref{IODS}). The red-times curve in Fig. \ref{fig:cosparse} shows the co-sparsity of the 1000 spikes using FOD matrix as the analysis dictionary. Based on FOD matrix, the MFOD matrix can be constructed as
\begin{equation}
\label{eq:Matrix_FODS}
\bm{\Omega}_\text{MFOD} \triangleq \frac{1}{\sqrt{q}}[\bm{D}^{f_0},\bm{D}^{f_1},\dots,\bm{D}^{f_{q-1}}]^T,
\end{equation}
with its order set
\begin{equation}
\bm{\mathcal{F}} = \left\{f_0,f_1,\dots,f_{q-1}\right\}\in\mathbb{R}^{q\times 1}.
\end{equation}
Define
\begin{equation}
d \triangleq \max|f_i-f_j|,\quad f_i,f_j\in\bm{\mathcal{F}},\quad i\neq j
\end{equation}
as the maximum order distance. It presents a trade-off between the co-sparsity of the analysis coefficients and the mutual coherence of $\bm{\Omega}_\text{MFOD}$. A small $d$ increases the co-sparsity of the analysis coefficients but leads to the condition that each $\bm{D}^i$ is highly coherent with the other order difference matrix, and vice versa. Our observations on different $d$ suggest that $d\in[\frac{1}{4},\frac{1}{2}]$ is a good compromise.

\textbf{Remark 1:} The redundant ratio $\rho$ presents a trade-off between signal reconstruction accuracy and computational cost. A large $\rho$ increases the reconstruction performance but consumes more computational resources. The work in \cite{nam2013cosparse} showed that $\rho\in[2,4]$ is a good compromise.

\section{Weighted Analysis $\ell_1$-Minimization}

\begin{figure*}[!htbp]
	\subfigure[]{\label{fig:coefficients}\includegraphics[width = .5\textwidth]{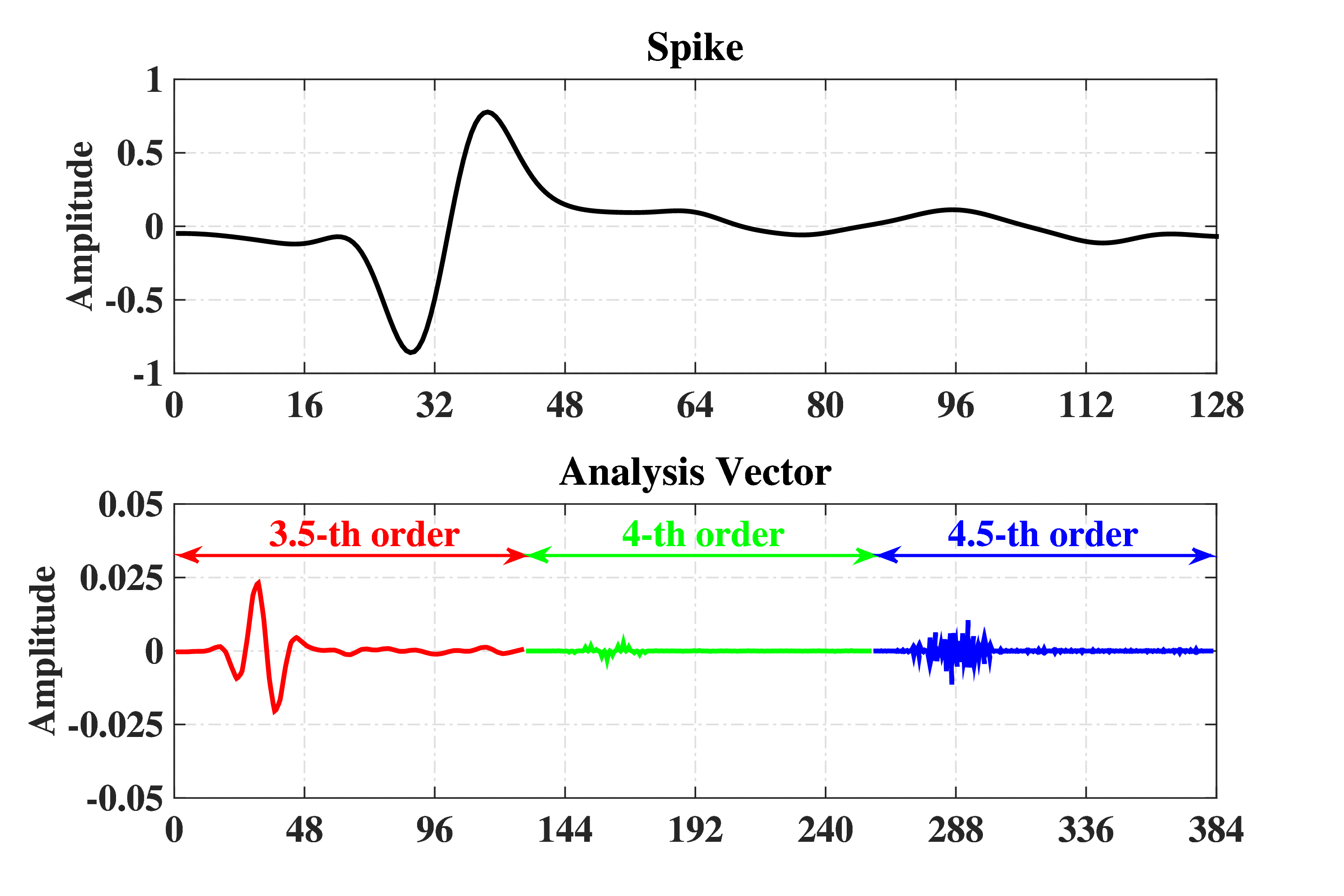}}
	\subfigure[]{\label{fig:histograms}\includegraphics[width = .5\textwidth]{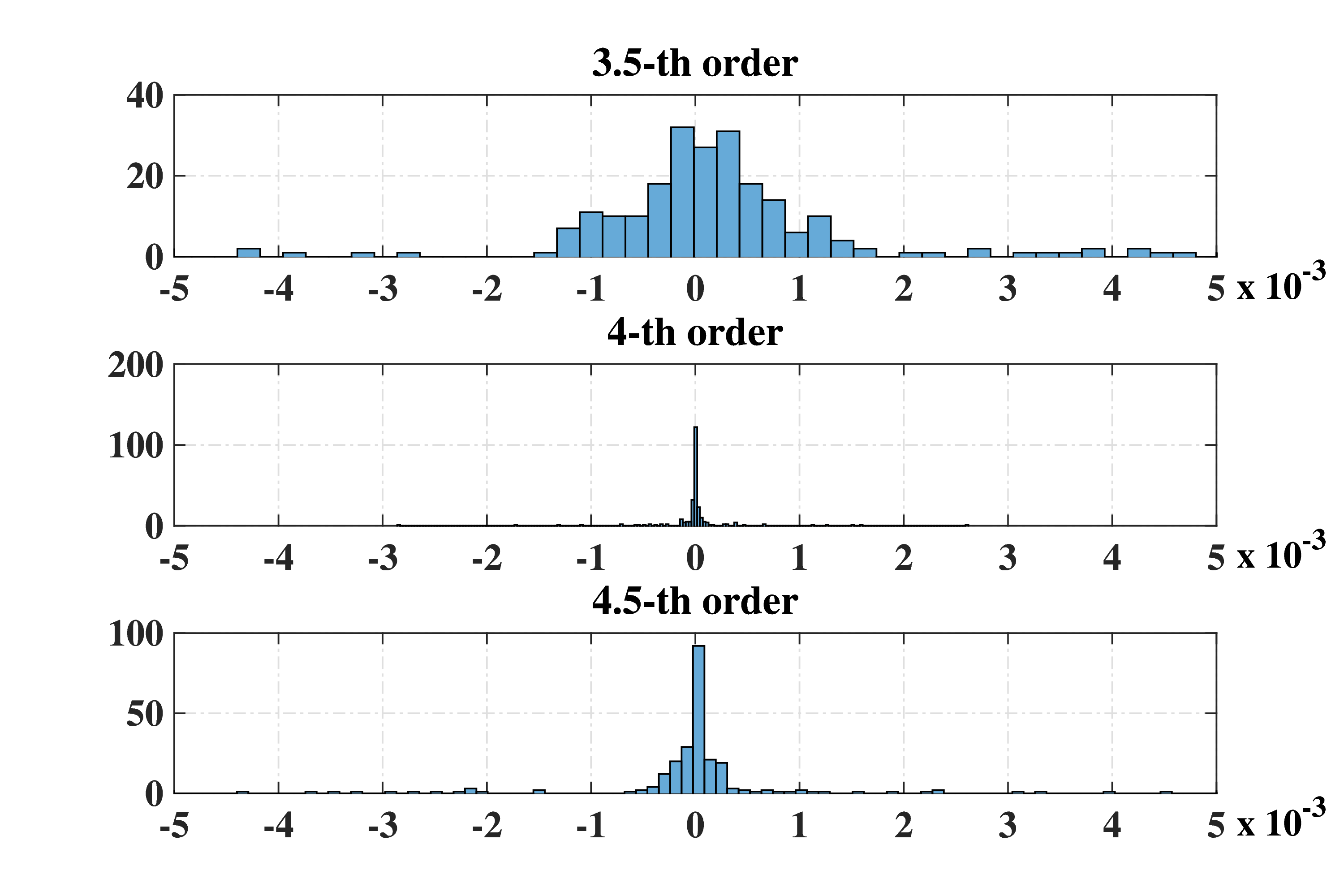}}
	\caption{(a) From top to bottom, the original neural spike and the corresponding analysis coefficients of 3.5-th, 4-th, and 4.5-th order difference. (b) From top to bottom, the histograms of analysis coefficients of the 3.5-th, 4-th, and 4.5-th order, respectively.}
	\label{fig12}
\end{figure*}

Having constructed the multiple fractional orders difference matrix $\bm{\Omega}$ as the analysis dictionary, we herein propose a weighted analysis $\ell_1$-minimization (WALM) method to reconstruct the original neural spike $\bm{x}$. Considering the CS measurement model $\bm{y} = \bm{\Phi x} + \bm{e}$, we assume that the components of $\bm{e}$ are independent and identically distributed (i.i.d.) Gaussian variables with unknown variance $\sigma_{\bm{e}}^2$, and the entry $z_i$ of $\bm{z} = \bm{\Omega}\bm{x}$ is independent and has a Laplacian distribution with standard deviation $\sigma_i$, i.e.,
\begin{equation}
p(z_i) = \frac{1}{\sqrt{2}\sigma_i}\text{exp}\left(-\frac{\sqrt{2}\|z_i\|_1}{\sigma_i}\right),\quad i = 1,\dots,l.
\end{equation}
Note that reconstructing $\bm{x}$ and $\bm{z}$ are identical. Therefore, we first infer $\bm{z}$ from $\bm{y}$ by maximizing the conditional probability distribution $p(\bm{z}|\bm{y},\bm{\Phi})$, which can be expressed by Bayes's rule as
\begin{equation}
p(\bm{z}|\bm{y},\bm{\Phi}) \propto p(\bm{y}|\bm{\Phi},\bm{z})p(\bm{z}).
\end{equation}
Because the noise $\bm{e}$ is assumed to be Gaussian, the likelihood function is given by
\begin{equation}
p(\bm{y}|\bm{\Phi},\bm{z}) \propto \text{exp}\left(-\frac{\|\bm{y}-\bm{\Phi}\bm{x}\|_2^2}{2\sigma_{\bm{e}}^2}\right).
\end{equation}
Hence, maximizing the posterior distribution $p(\bm{z}|\bm{y},\bm{\Phi})$ leads to
\begin{equation}
\label{eq:MAPz}
\begin{split}
\hat{\bm{z}}_\text{MAP} &= \arg\underset{\bm{z}}{\max}\ p(\bm{z}|\bm{y},\bm{\Phi})\\
&= \arg\underset{\bm{z}}{\max}\left(\log p(\bm{y}|\bm{\Phi},\bm{z})+\sum_i\log p(z_i)\right).\\
\end{split}
\end{equation}
By substituting $\bm{z} = \bm{\Omega}\bm{x}$ into (\ref{eq:MAPz}), we obtain
\begin{equation}
\label{eq:MAPx}
\hat{\bm{x}}_\text{MAP} = \arg\underset{\bm{x}}{\min}\left(\frac{\|\bm{y}-\bm{\Phi}\bm{x}\|_2^2}{2\sigma_{\bm{e}}^2}+\sum_i\frac{\sqrt{2}\|\bm{\Omega}_i \bm{x}\|_1}{\sigma_i}\right),
\end{equation}
where $\bm{\Omega}_i$ denotes the $i$th row of $\bm{\Omega}$, $i\in\{1,\dots,p\}$. The problem in (\ref{eq:MAPx}) is equivalent to
\begin{equation}
\label{eq:xMAP}
\hat{\bm{x}}_\text{MAP} = \arg\underset{\bm{x}}{\min}\frac{1}{2}\|\bm{y}-\bm{\Phi}\bm{x}\|_2^2+\lambda\|{\rm diag}(\bm{w})\bm{\Omega}\bm{x}\|_1,
\end{equation}
where $\bm{w} = [\frac{1}{\sigma_1},\frac{1}{\sigma_2},\dots,\frac{1}{\sigma_l}]$ and $\lambda$ denotes a tuning parameter. Hence, the $\ell_1$-minimization in (\ref{eq:analysisl1}) can be interpreted as the MAP estimation under the hypothesis that all $\sigma_i$ are equal.

However, for our multiple fractional orders analysis matrix $\bm{\Omega}$, the hypothesis that the entries of $\bm{z}$ have equal standard deviations does not reflect this fact. To illustrate this argument, Fig. \ref{fig:coefficients} plots the analysis coefficients of the 3.5-th, 4-th, and 4.5-th order difference for a typical neural spike\footnote{The spike was randomly chosen from the Leicester Easy2 dataset \cite{quiroga2004unsupervised}.}. At the same time, the corresponding histograms of the analysis coefficients in the three orders are simultaneously reported in Fig. \ref{fig:histograms}.

\begin{figure}[!htbp]
	\centering
	\includegraphics[width = .5\textwidth]{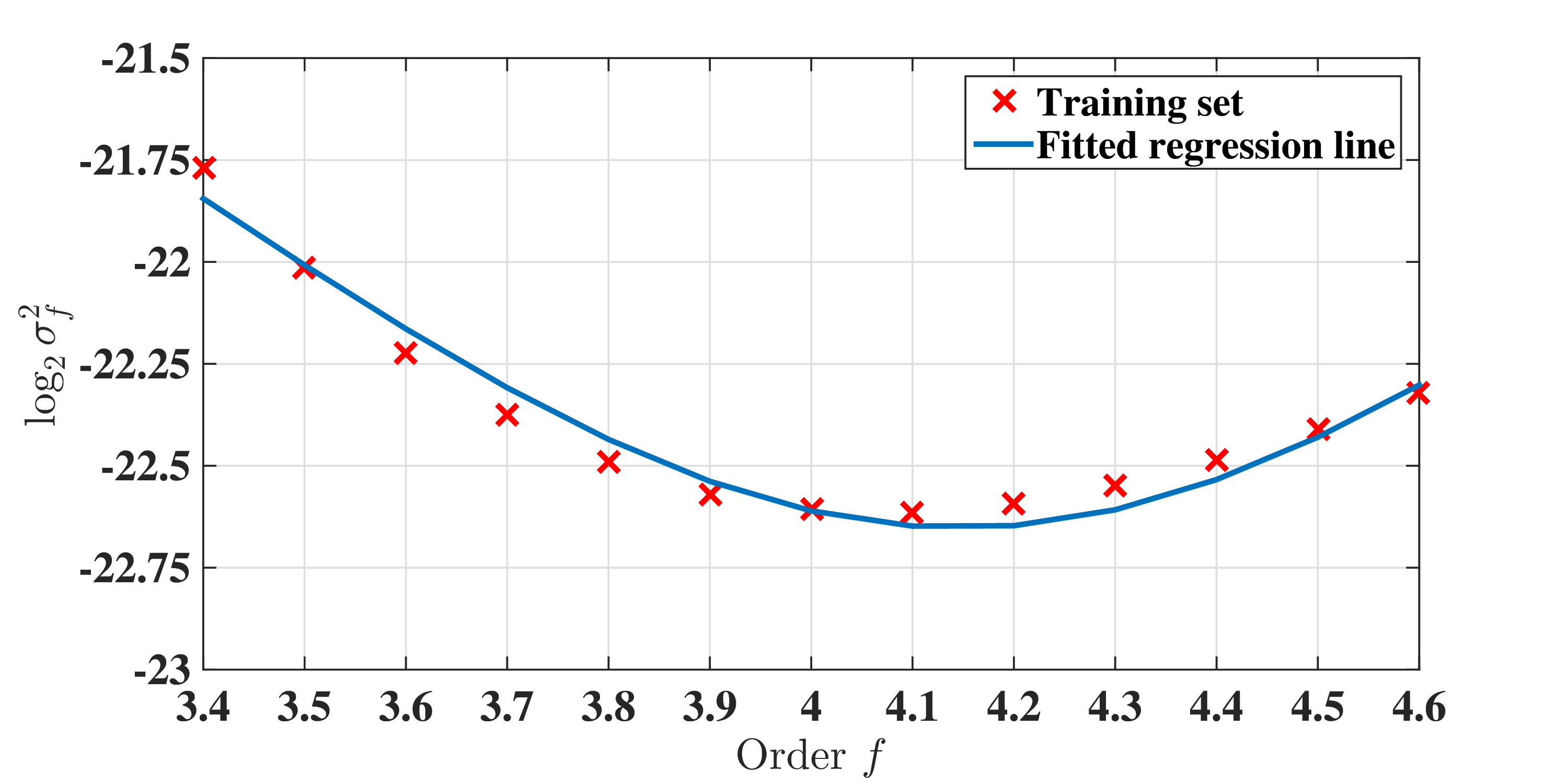}
	\caption{Scatter plot of fractional order $f$ versus variance $\log_2\sigma_f^2$ and fitted regression curve. For each order, the variance was averaged over the training dataset that contains 100 spikes randomly chosen from Leicester Easy2 dataset.}
	\label{fig:regression}
\end{figure}

Clearly, the standard deviations of analysis coefficients of distinct orders are not identical. To cope with this issue, we divide the standard deviation vector $\bm{w}$ into multiple groups to incorporate the aforementioned multiple orders prior. Suppose $\bm{\Omega}$ is constructed of $q$ difference matrices with fractional orders, then $\bm{w}$ can be partitioned into $q$ groups, i.e.,
\begin{equation}
\label{eq:multi-w}
\bm{w} = \left[\bm{w}^T_{G_0},\bm{w}^T_{G_1},\dots,\bm{w}^T_{G_{q-1}}\right]^T,
\end{equation}
where $\bm{w}^T_{G_0},\bm{w}^T_{G_1},\dots,\bm{w}^T_{G_{q-1}}$ represent the standard deviations corresponding to $q$ orders, respectively. Note that all $\sigma_i$ of the same group are equal. As the variance of analysis coefficients tends to decrease first and increase across orders, we propose to model the variance across orders with quadratic functions as
\begin{equation}
\label{eq:prior}
\sigma_{f_i}^2 = c_i2^{-2a_if_i^2 - 2b_if_i},\quad i = 0,\dots,q-1,
\end{equation}
where $a_i$, $b_i$ and $c_i$ are the model parameters, and $f_i$ is the difference order. In this model, the $\sigma_i$ are made equal for all coefficients within an order, and $\sigma^2_{f_i}$ refers to the variance of the analysis coefficients at order $f_i$. Therefore, we have
\begin{equation}
\label{eq:w}
\bm{w}_{G_i} = \sigma_{f_i} = \frac{2^{a_if_i^2+b_if_i}}{\sqrt{c_i}},\quad i = 0,\dots,q-1.
\end{equation}
As the entries of $\bm{w}$ only depend on the value of $\bm{a}$, $\bm{b}$, and $\bm{c}$, problem (\ref{eq:xMAP}) can be solved after these parameters are calculated. This leads us to propose a training stage to estimate these values. The first part predicts the standard deviations $\sigma_{f_i}$ using maximum likelihood estimation. Once the variances are estimated, simple quadratic regression can be employed to solve for $a_i$, $b_i$, and $c_i$ in the following equation, derived from (\ref{eq:prior}),
\begin{equation}
\label{eq:sigma}
\log_2\sigma_{f_i}^2 = \log_2c_i-2a_if_i^2-2b_if_i.
\end{equation}
An example of the fitted regression curve is shown in Fig. \ref{fig:regression}.

After building $\bm{w}$, the problem (\ref{eq:xMAP}) can be easily solved using $\ell_1$-minimization algorithms. The complete WALM algorithm can be outlined as Algorithm \ref{alg:WALM}.

\textbf{Remark 2:} The proposed WALM is a nearly signal independent approach. Although it requires a training step to estimate the regression parameters, the amount of data needed is much less than that of signal dependent approaches such as \cite{zhang2014efficient} and \cite{suo2014energy}. Therefore, WALM can significantly reduce the data storage and computational resource.

\textbf{Remark 3:} In contrast to the canonical AL1 method, the main advantage of WALM is the incorporation of the multiple orders prior in analysis coefficients, including the positions of nonzero coefficients and its standard deviations between neighboring difference orders, which will allow the number of measurements to be significantly reduced without leading to ambiguity.

\begin{algorithm}[htbp]
	\caption{Weighted Analysis $\ell_1$-Minimization}
	\label{alg:WALM}
	\begin{algorithmic}[1]
		\Require
		$\bm{y},\bm{\Phi},q$
		\For{$i=1,\dots,q-1$}
			\State Construct $\bm{D}^{f_i}$ by using (\ref{eq:FODS})
		\EndFor
		\State Construct $\bm{\Omega}$ by using (\ref{eq:Matrix_FODS})
		\For{$i=1,\dots,q-1$}
			\State Estimate $a_i$, $b_i$, and $c_i$ by using (\ref{eq:sigma})
			\State Construct $\bm{w}^T_{G_i}$ by using (\ref{eq:w})
		\EndFor
		\State Construct $\bm{w}$ by using (\ref{eq:multi-w})
		\State Solve $\hat{\bm{x}} = \arg\underset{\bm{x}}{\min}\frac{1}{2}\|\bm{y}-\bm{\Phi}\bm{x}\|_2^2+\lambda\|{\rm diag}(\bm{w})\bm{\Omega}\bm{x}\|_1$
		\Ensure
		recovered signal $\hat{\bm{x}}$
	\end{algorithmic}
\end{algorithm}

\section{Experiment Validation}
In this section, we examine the performance of the proposed WALM method against state-of-the-art compressed sensing schemes for implant neural recording.

\subsection{Experimental Setup}
We use the Leicester neural signal database \cite{quiroga2004unsupervised}, which contains 20 synthesized datasets. Each dataset contains spikes from three different neurons with different noise levels. The datasets are categorized by the spike sorting difficulty levels, such as Leicester Difficult1, Difficult2, Easy1, and Easy2. We take 128 samples around each spike to form the signal frame. To simplify the comparison, we retain the signal containing only one spike. A Random i.i.d Bernoulli matrix is chosen as the sensing matrix because it guarantees excellent reconstruction quality and implementation efficiency \cite{chen2012design}. All signals are compressed and reconstructed for 20 times, using a different sensing matrix in each trial. The results are then averaged across all trials. To measure the reconstruction quality, we employ the percentage root-mean-square difference (PRD) to quantify the error percentage between the original $\bm{x}$ and the reconstructed $\hat{\bm{x}}$:
\begin{equation}
\text{PRD} = \frac{\|\bm{x}-\hat{\bm{x}}\|_2}{\|\bm{x}\|_2}\times 100\%.
\end{equation}
For physiological signal reconstruction, Zigel \emph{et al.} \cite{zigel2000weighted} classified the different values of PRD based on the signal quality perceived by specialists. In this work, PRD value below $5\%$ is regarded as ``good'' reconstruction quality.

The following state-of-the-art CS algorithms have been chosen for performance comparison.

1) Basis Pursuit De-noising (BPDN) described in (\ref{eq:l1}). The orthonormal basis of Daubechies-4 wavelet was used as the sparsity dictionary. For the BPDN implementation, we used the solvers \emph{SolveBP} from the SparseLab toolbox \cite{donoho2008sparselab}.

2) Block Sparse Bayesian Learning (BSBL) proposed by Zhang \emph{et al.} \cite{zhang2013extension}. We used the solver \emph{BSBL-BO} for BSBL implementation.

3) Analysis $\ell_1$-minimization (AL1) algorithm described in (\ref{eq:analysisl1}). For the AL1 implementation, we used the \emph{CVX} toolbox \cite{grant2008cvx} from Stanford University.

4) Signal Dependent Neural Compressed Sensing (SDNCS) method proposed in \cite{zhang2014efficient}. It used a sparse representation dictionary learned from data. For this method, each dataset was divided into training section and test section, composed of $20\%$ and $80\%$ of the dataset. The training section was used to construct the sparse representation dictionary, whereas the test section was used to evaluate its performance\footnote{The implementation of SDNCS can be downloaded from http://etienne.ece.jhu.edu/projects/neural\_cs/CS\_code.rar}.

\subsection{The Advantage of MFOD Matrix}
\begin{figure}[tbp]
\centering
\subfigure[]{\label{fig:Exp4a}\includegraphics[width = .5\textwidth]{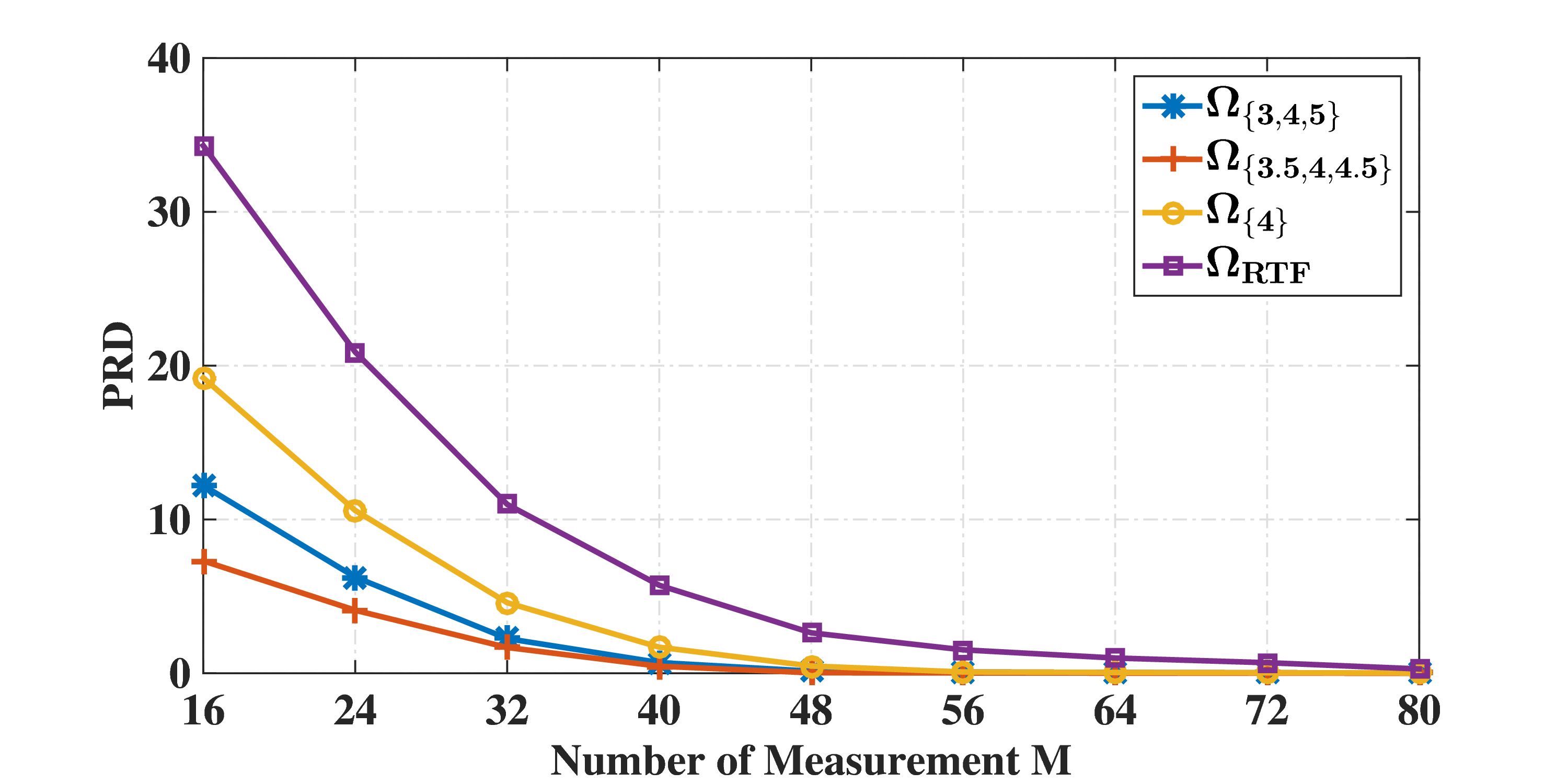}}
\subfigure[]{\label{fig:Exp4b}\includegraphics[width = .5\textwidth]{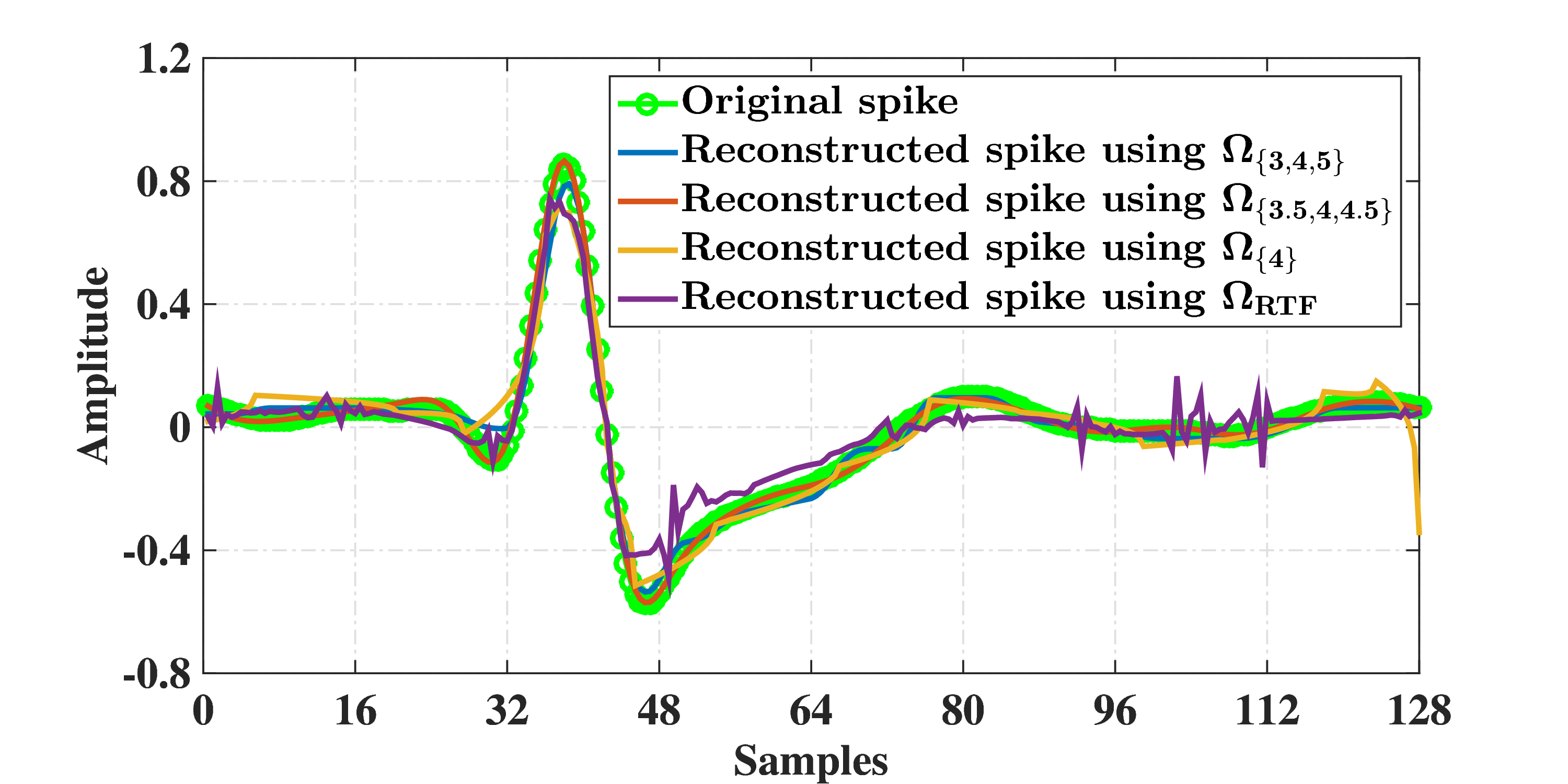}}
\caption{(a) Averaged PRDs over all spikes from Leicester Easy1 dataset versus the different number of measurements $M$ for $\bm{\Omega_{\{3,4,5\}}}$, $\bm{\Omega_{\{3.5,4,4.5\}}}$, $\bm{\Omega_{\{4\}}}$, and $\bm{\Omega_{\text{RTF}}}$, respectively. (b) Original spike and reconstructed spikes using $\bm{\Omega_{\{3,4,5\}}}$, $\bm{\Omega_{\{3.5,4,4.5\}}}$, $\bm{\Omega_{\{4\}}}$, and $\bm{\Omega_{\text{RTF}}}$, respectively.}
\label{fig:Exp4}
\end{figure}

To evaluate the effectiveness of the proposed MFOD matrix, we compared it with IOD matrix, MIOD matrix, and the random tight frame (RTF) proposed in \cite{nam2013cosparse}. In this experiment, we constructed IOD matrix $\bm{\Omega_{\{4\}}}$ as (\ref{IODS}), MIOD matrix $\bm{\Omega_{\{3,4,5\}}}$ as (\ref{eq:RAD}), and MFOD matrix $\bm{\Omega_{\{3.5,4,4.5\}}}$ as (\ref{eq:Matrix_FODS}). The Leicester Easy1 dataset was chosen for evaluation. For all the four dictionaries, the AL1 algorithm was used to reconstruct the spikes. The average PRDs over all spikes for the four dictionaries and a spike reconstruction example are shown in Fig. \ref{fig:Exp4a} and \ref{fig:Exp4b}, respectively. Among the four different dictionaries, the $\bm{\Omega_\text{RTF}}$ has the worst performance. It is mainly because the $\bm{\Omega_\text{RTF}}$ is a general analysis dictionary and does not exploit any statistical information of neural spikes. Furthermore, we can note that $\bm{\Omega_{\{3.5,4,4.5\}}}$ and $\bm{\Omega_{\{3,4,5\}}}$ outperform $\bm{\Omega_{\{4\}}}$ due to their redundancy. In addition, $\bm{\Omega_{\{3.5,4,4.5\}}}$ has better reconstruction accuracy than $\bm{\Omega_{\{3,4,5\}}}$, especially when the number of measurements is very small. It is mainly because the analysis coefficients with $\bm{\Omega_{\{3.5,4,4.5\}}}$ will be sparser than that of $\bm{\Omega_{\{3,4,5\}}}$, and the high sparsity reduces the number of measurements for signal reconstruction.

\subsection{Average PRD and the Probability of ``Good'' Reconstruction}
\begin{figure}[tbp]
\centering
\subfigure[]{\label{fig:Exp1a}\includegraphics[width = .5\textwidth]{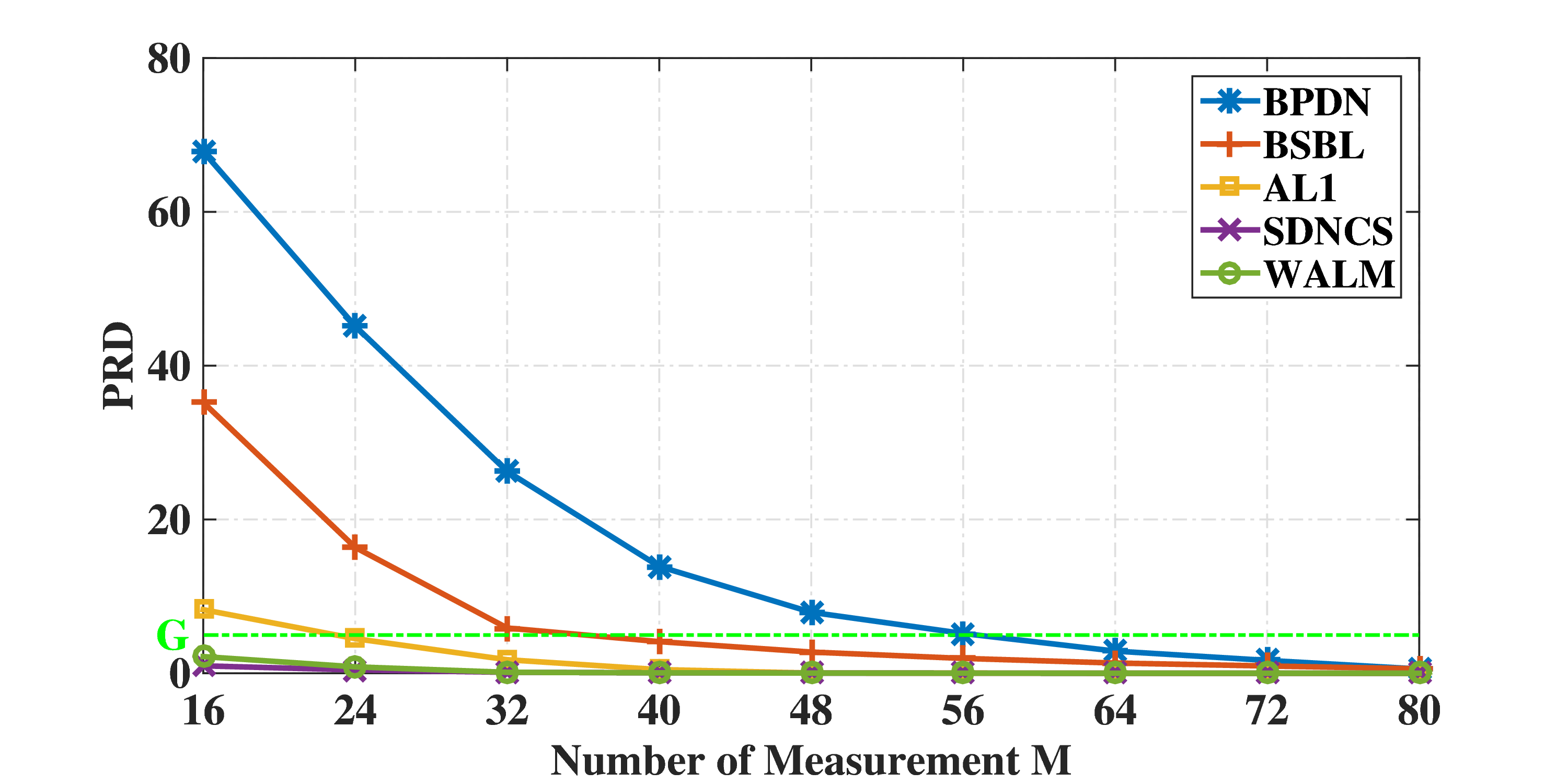}}
\subfigure[]{\label{fig:Exp1b}\includegraphics[width = .5\textwidth]{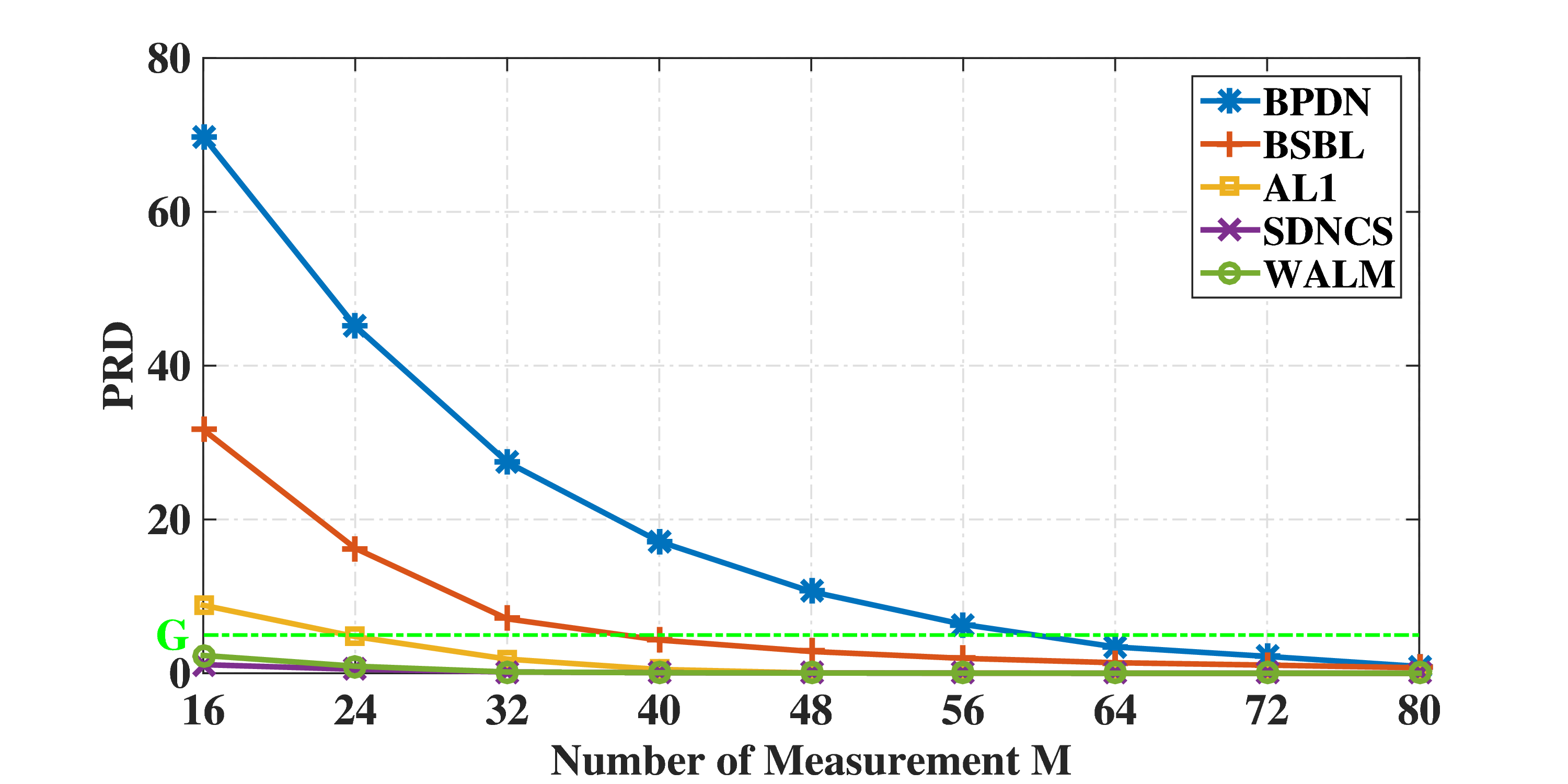}}
\caption{PRD averaged over all spikes from (a) Easy1 dataset, (b) Difficult1 dataset, versus the different number of measurements $M$ for BPDN, BSBL, AL1, SDNCS, and WALM, respectively. The green dash-dotted line denotes the ``good'' PRD bound at $5\%$.}
\label{fig:Exp1}
\end{figure}

Then we evaluate the performance of the proposed WALM algorithm versus the number of measurements. The $\bm{\Omega_{\{3.5,4,4.5\}}}$ was used as the analysis dictionary for both AL1 and WALM. For the Leicester neural signal database, the Easy1 and Difficult1 datasets are chosen for evaluation. The experimental results are shown in Fig. \ref{fig:Exp1}, where each point indicates the average PRD of all spikes at a specified number of measurements. At the same time, Table \ref{tab:good} reports the probability of ``good'' reconstruction quality in different situations. First of all, we can observe that analysis model based algorithms outperform synthesis model based ones in terms of both averaged PRD and the probability of ``good'' reconstruction quality. Moreover, due to the incorporation of the multiple orders prior in analysis coefficients, the WALM algorithm performs better than the canonical AL1 algorithm, especially when the number of measurements is very small. The WALM algorithm has the averaged PRD less than $5\%$ for all numbers of measurements, and it achieves more than $92\%$ of ``good'' reconstruction quality with $M=16$. As a comparison, BPDN, BSBL, and AL1 cannot recover so many spikes both in Easy1 and Difficult1 datasets with ``good'' reconstruction quality under this condition. Both SDNCS and WALM algorithms show ``good'' reconstruction quality (SDNCS even slightly outperforms WALM when $M=16$), while the proposed WALM algorithm simplifies sparse dictionary learning with much fewer computational resources, which is preferred for practical neural recording experiments.

\begin{table*}[tbp]
\footnotesize
\centering
\doublerulesep=1pt
\tabcolsep=2pt
\caption{\label{tab:test} Probabilities of Reconstruction for ``good'' Quality Under the Different Number of Measurements $M$ ($\%$).}
\label{tab:good}
\begin{tabular}{ccccccccccccc}
	\hline
	\hline
	&& \multicolumn{5}{c}{Easy1 dataset} && \multicolumn{5}{c}{Difficult1 dataset} \\
	\cline{3-7}\cline{9-13}
	\raisebox{1ex}[0pt]{$M$} && BPDN & BSBL & AL1 & SDNCS & WALM && BPDN & BSBL & AL1 & SDNCS & WALM \\
	\hline
	16 && 0 & 0 & 30.8 & 93.3 & 92.5 && 0 & 0 & 28.8 & 93.1 & 92.2 \\
	24 && 0 & 2.1 & 54.2 & 95.6 & 95.1 && 0 & 1.3 & 52.1 & 94.9 & 94.3 \\
	32 && 0 & 52.9 & 83.8 & 97.9 & 97.7 && 0 & 48.5 & 82.4 & 96.2 & 96.3 \\
	40 && 9.3 & 60.8 & 92.1 & 100 & 100 && 6.4 & 58.3 & 90.2 & 98.8 & 98.9 \\
	48 && 42.9 & 68.4 & 94.3 & 100 & 100 && 33.4 & 66.7 & 93.0 & 100 & 100 \\
	56 && 52.9 & 75.8 & 98.8 & 100 & 100 && 48.2 & 74.3 & 98.7 & 100 & 100 \\
	64 && 68.2 & 80.8 & 100 & 100 & 100 && 63.3 & 78.4 & 100 & 100 & 100 \\
	72 && 84.5 & 93.2 & 100 & 100 & 100 && 82.8 & 92.0 & 100 & 100 & 100 \\
	80 && 92.7 & 98.3 & 100 & 100 & 100 && 91.2 & 97.9 & 100 & 100 & 100 \\
	\hline
	\hline
\end{tabular}
\end{table*}

\begin{figure}[htb]
	\subfigure[]{\label{fig:Exp2a}\includegraphics[width = .5\textwidth]{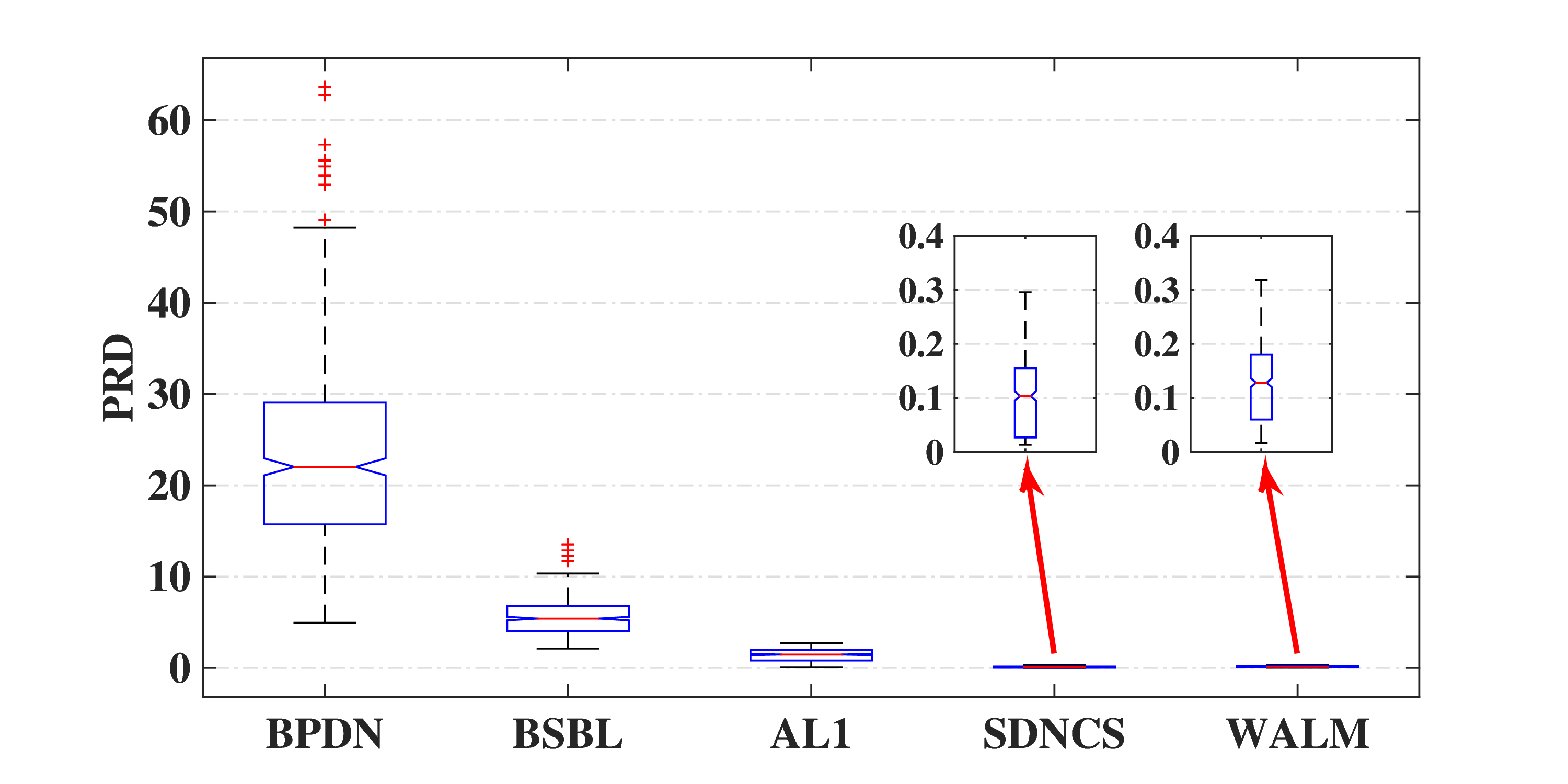}}
	\subfigure[]{\label{fig:Exp2b}\includegraphics[width = .5\textwidth]{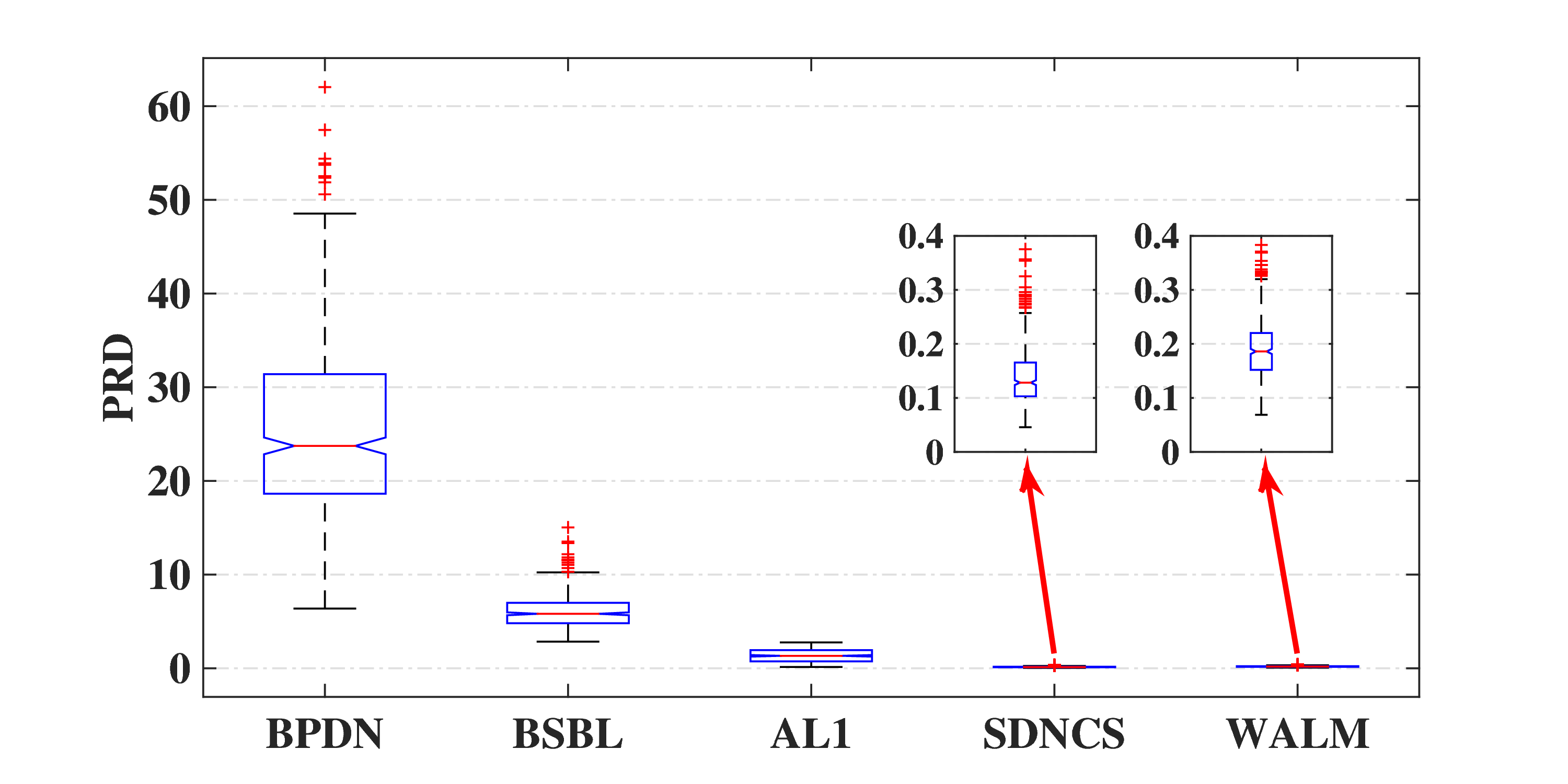}}
	\caption{Box plots for all spikes from (a) Easy1 dataset, (b) Difficult1 dataset, for BPDN, BSBL, AL1, SDNCS, and WALM, respectively, when the number of measurements is $M = 32$.}
	\label{fig:Exp2}
\end{figure}

To observe the PRD variance across individual datasets, Fig. \ref{fig:Exp2} shows the box plots for these algorithms when the number of measurements is $M = 32$. On each box, the central mark indicates the median, the edges of the box are the 25th and 75th percentiles, and the whiskers extend to the most extreme data points. Obviously, both for Easy1 and Difficult1 datasets, the WALM algorithm adopted the multiple fractional orders analysis dictionary outperforms the other algorithms. Once more, SDNCS has similar performance as WALM. Moreover, although the PRD variances of the two datasets are similar, the number of outliers of Difficult1 dataset is more than that of Easy1 dataset.

\subsection{Performance of Classification using Reconstructed Spikes}
To further illustrate the performance of WALM, we carried out a spike classification experiment using reconstructed spikes. The Leicester Easy1 and Difficult1 datasets were chosen for evaluation. Firstly, all spikes were compressed with $M=16$ and reconstructed using the five algorithms. Then Principal component analysis (PCA) and \cite{jolliffe2002principal} wavelet decomposition \cite{quiroga2004unsupervised} methods were used to extract features from reconstructed spikes in Easy1 and Difficult1 datasets, respectively. Finally, the first 10 features of each spike were used for classification by superparamagnetic clustering (SPC) \cite{quiroga2004unsupervised} algorithm.

\begin{figure*}[tbp]
	\centering
	\subfigure[]{\label{fig:Exp5a}\includegraphics[width = .3\textwidth]{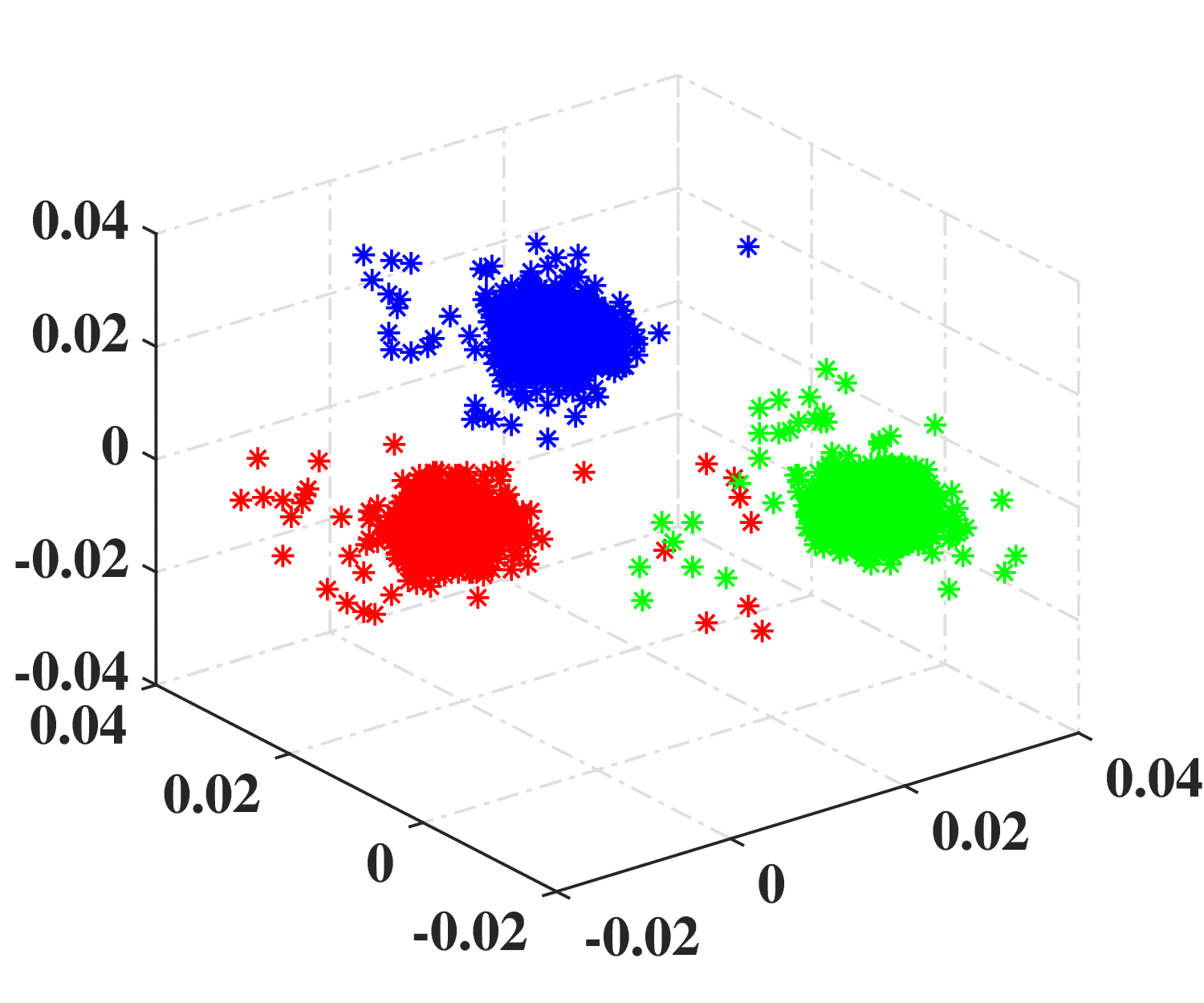}}
	\subfigure[]{\label{fig:Exp5b}\includegraphics[width = .3\textwidth]{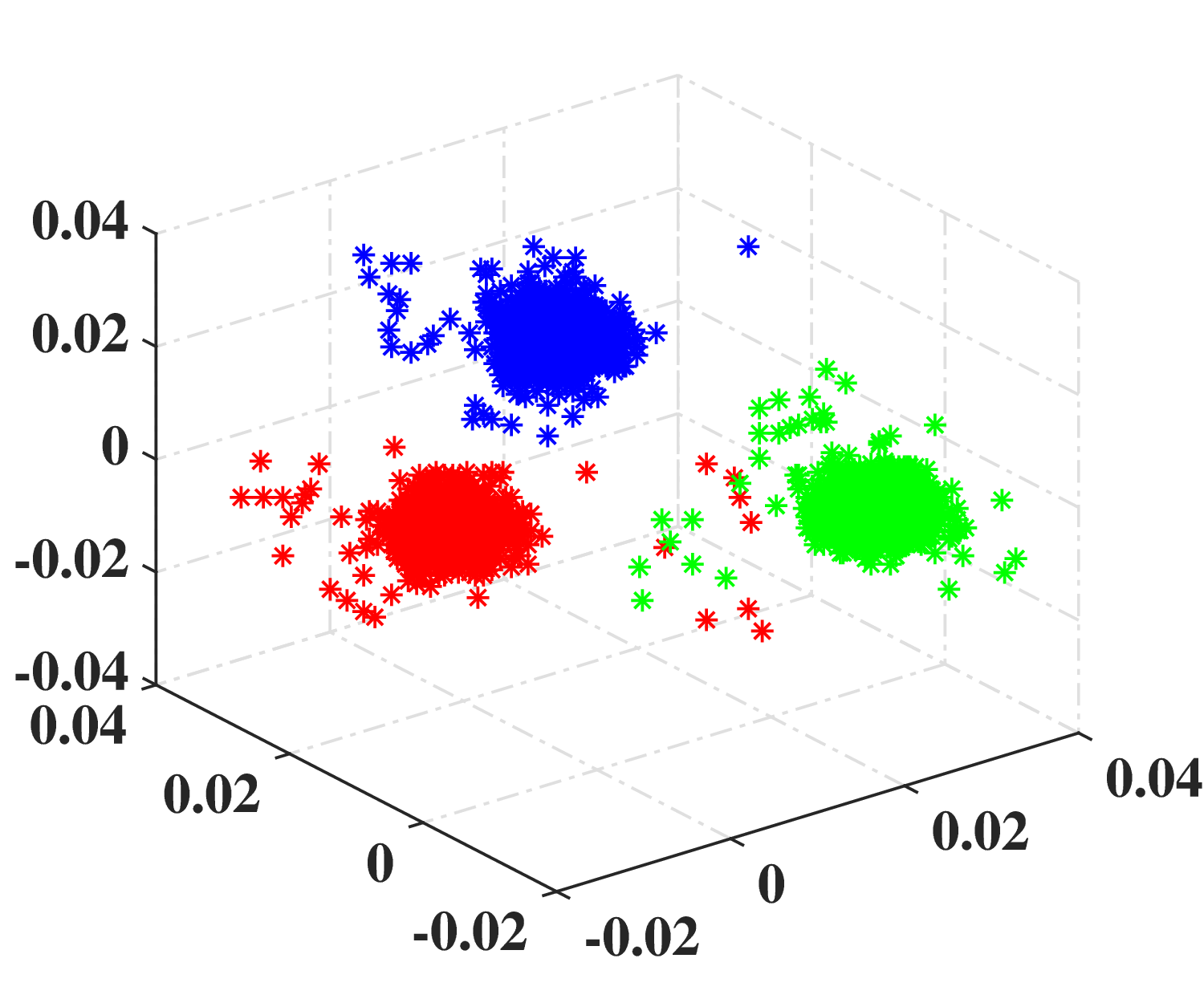}}
	\subfigure[]{\label{fig:Exp5c}\includegraphics[width = .3\textwidth]{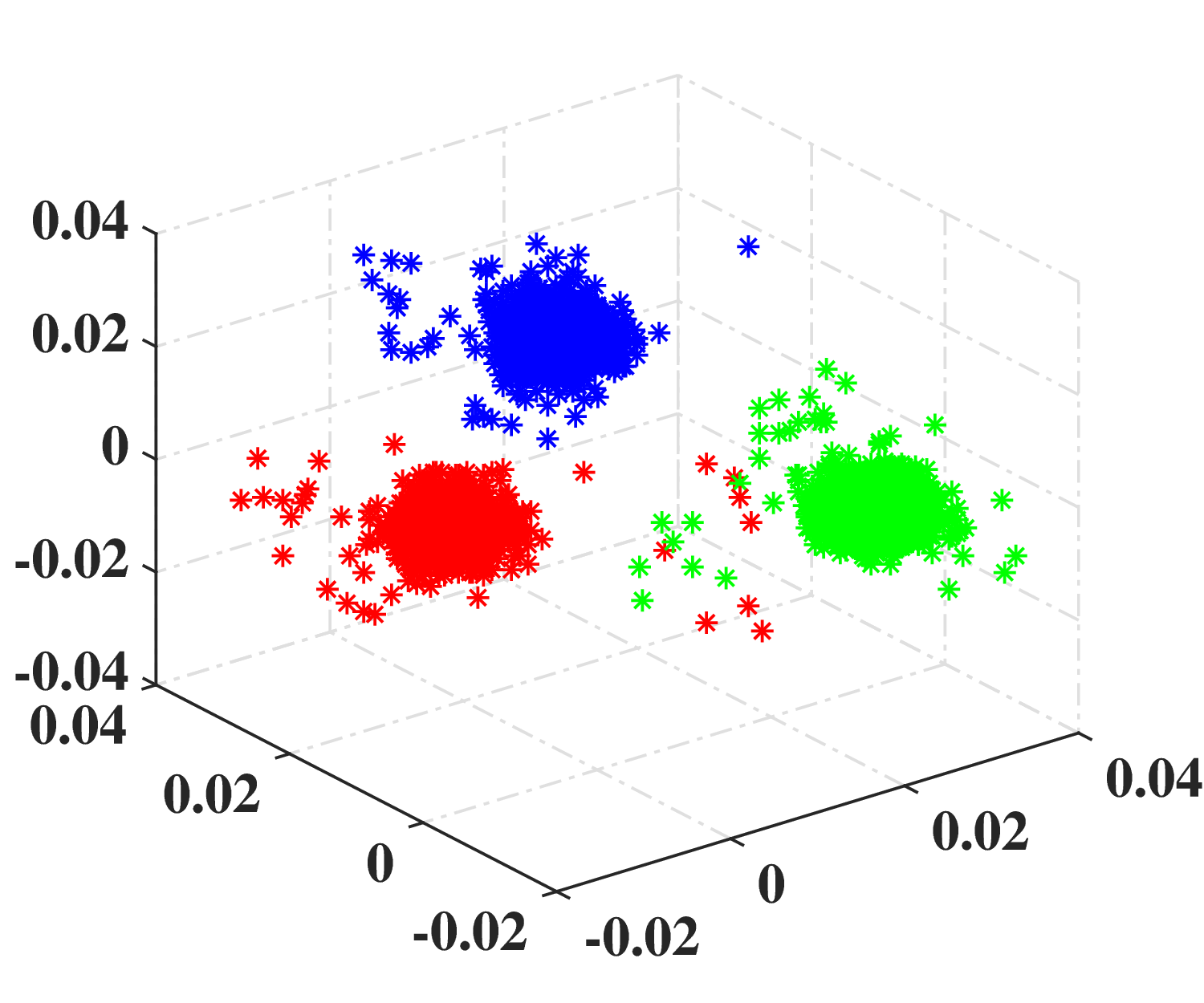}}
	\subfigure[]{\label{fig:Exp5d}\includegraphics[width = .3\textwidth]{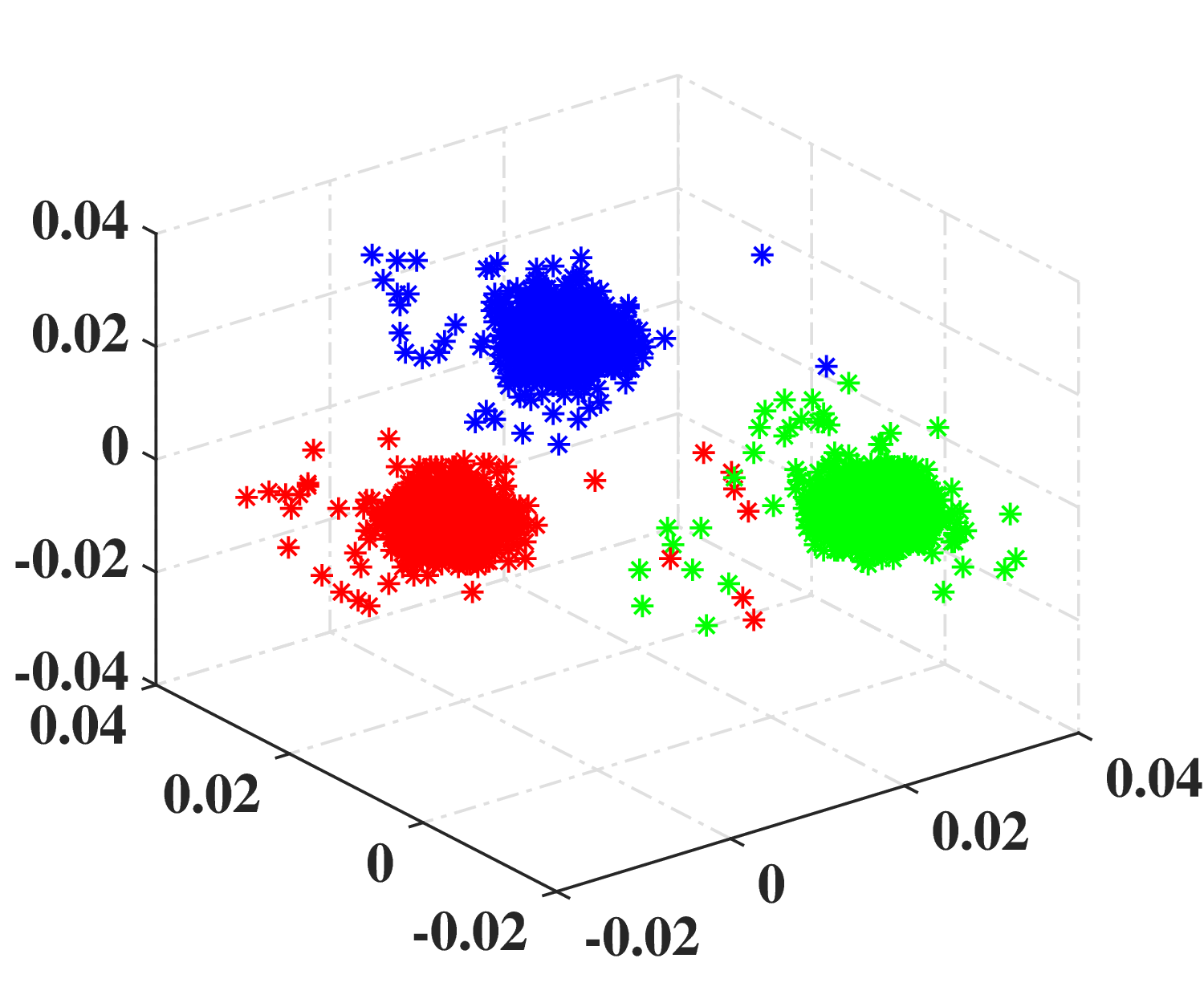}}
	\subfigure[]{\label{fig:Exp5e}\includegraphics[width = .3\textwidth]{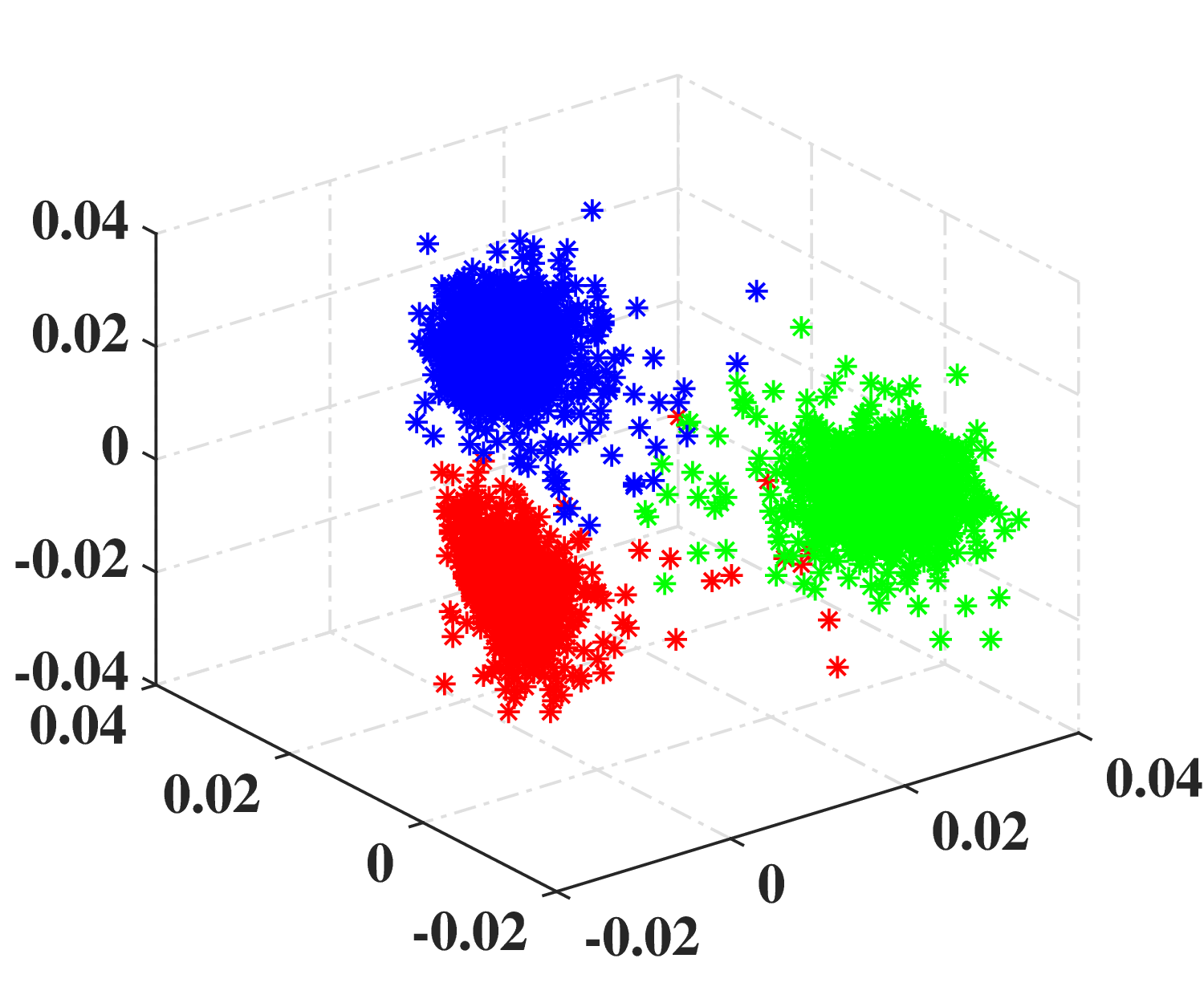}}
	\subfigure[]{\label{fig:Exp5f}\includegraphics[width = .3\textwidth]{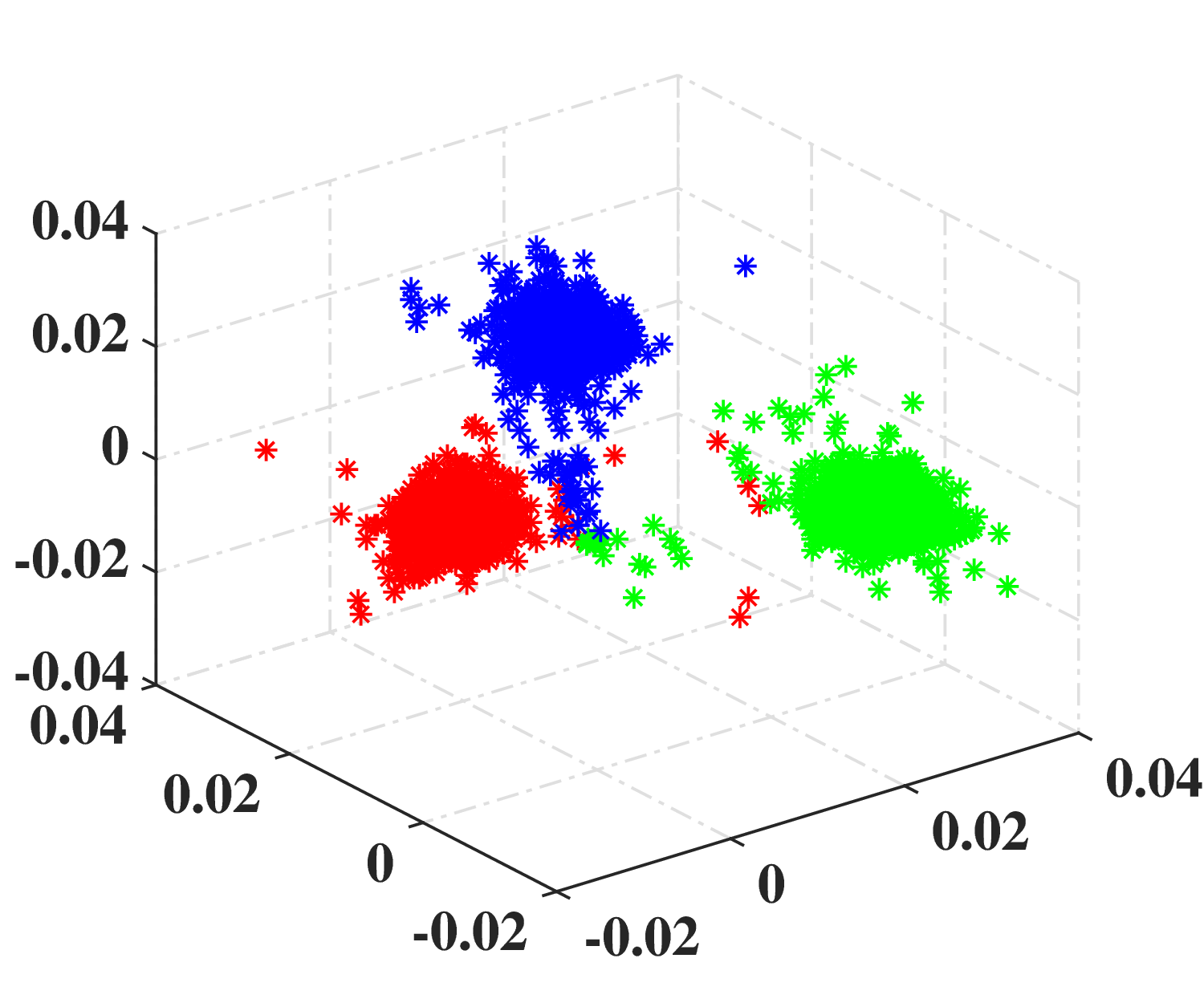}}
	\caption{Scatter plot of the first three features of (a) original spikes in Easy1 dataset, (b) spikes reconstructed by WALM, (c) spikes reconstructed by SDNCS, (d) spikes reconstructed by AL1, (e) spikes reconstructed by BPDN, and (f) spikes reconstructed by BSBL, respectively. The features were extracted using PCA. Points are colored according to the cluster to which they are assigned. The number of measurements is $M = 16$.}
	\label{fig:Exp5}
\end{figure*}

\begin{figure*}[tbp]
	\centering
	\subfigure[]{\label{fig:Exp6a}\includegraphics[width = .3\textwidth]{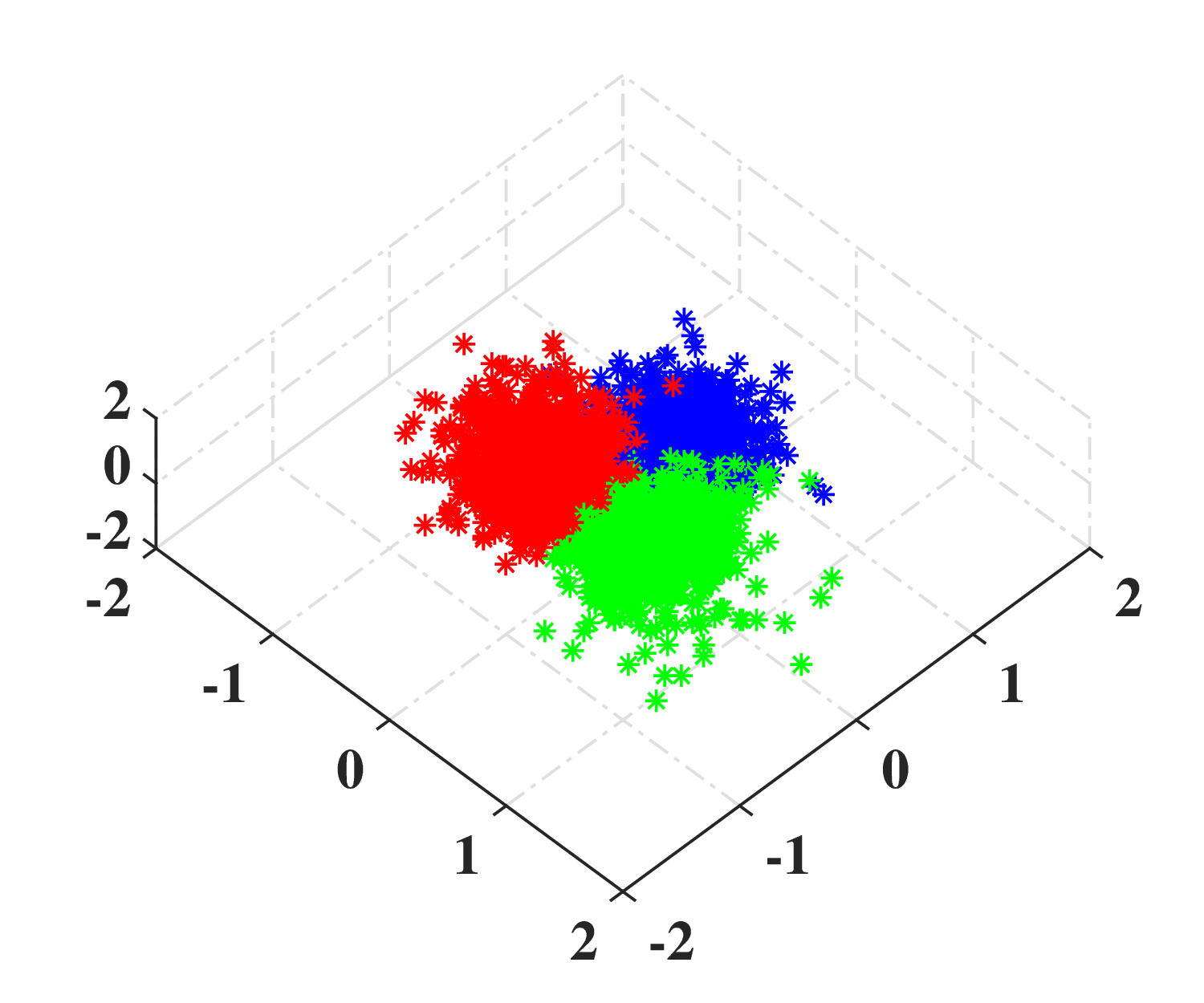}}
	\subfigure[]{\label{fig:Exp6b}\includegraphics[width = .3\textwidth]{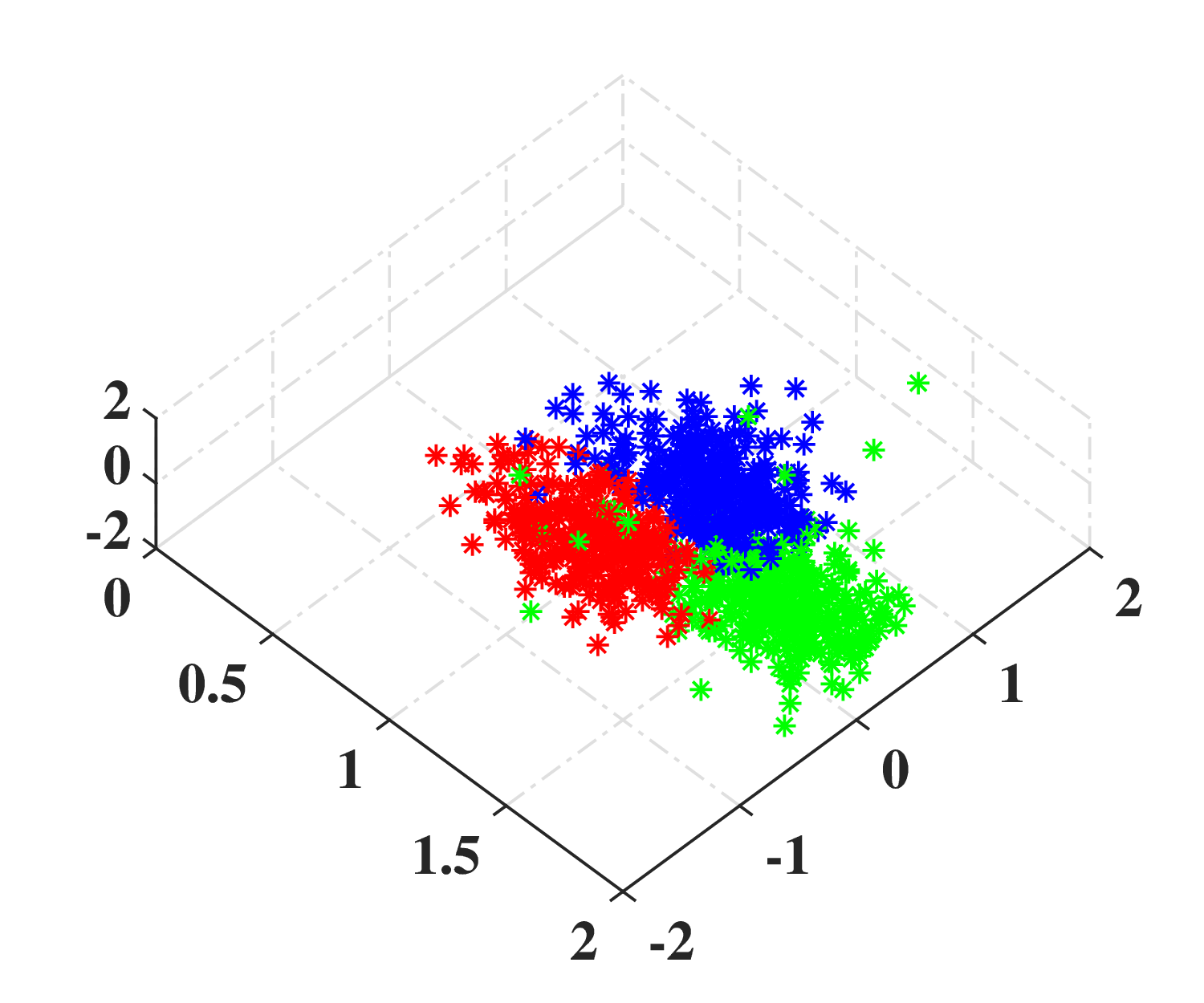}}
	\subfigure[]{\label{fig:Exp6c}\includegraphics[width = .3\textwidth]{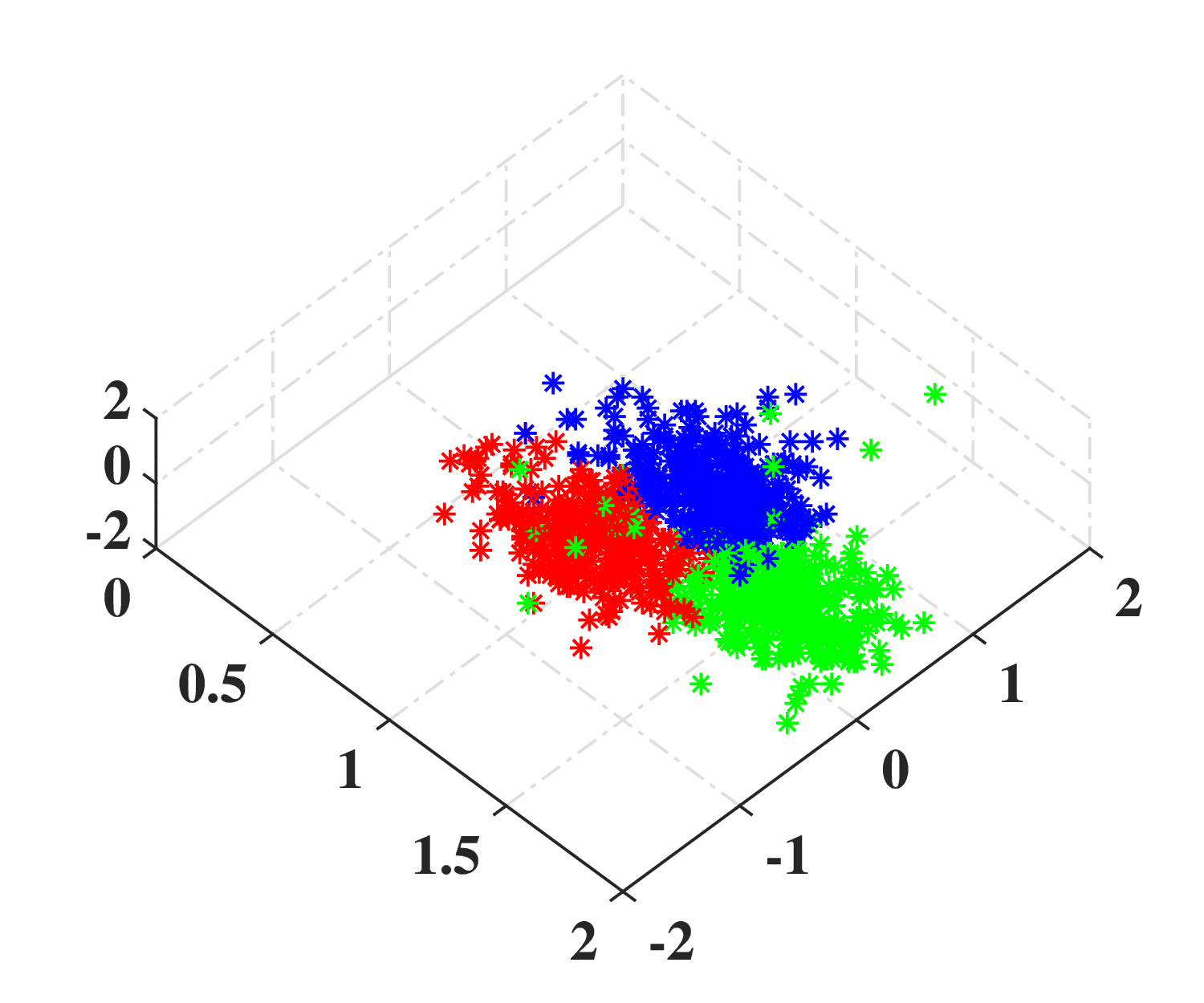}}
	\subfigure[]{\label{fig:Exp6d}\includegraphics[width = .3\textwidth]{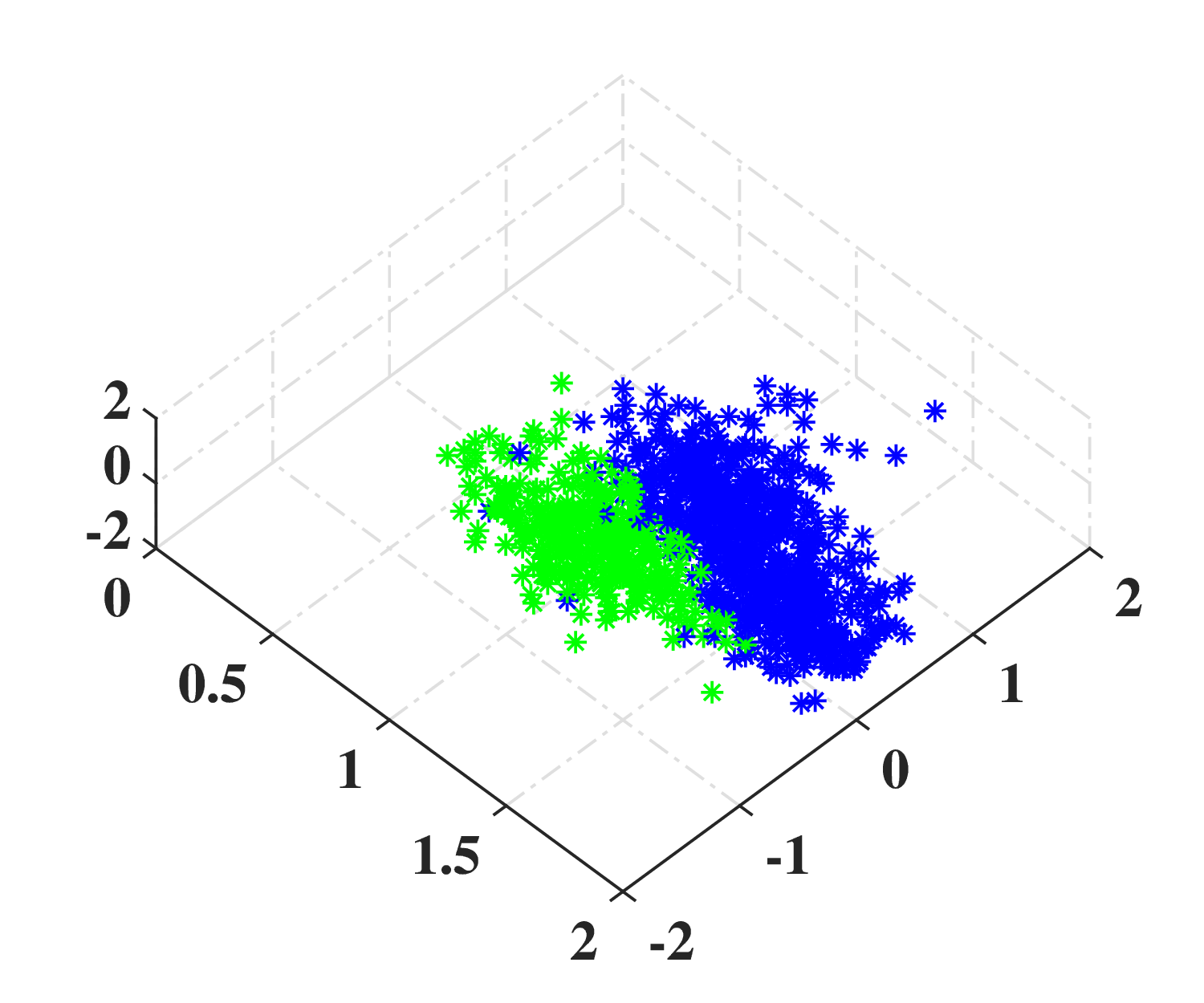}}
	\subfigure[]{\label{fig:Exp6e}\includegraphics[width = .3\textwidth]{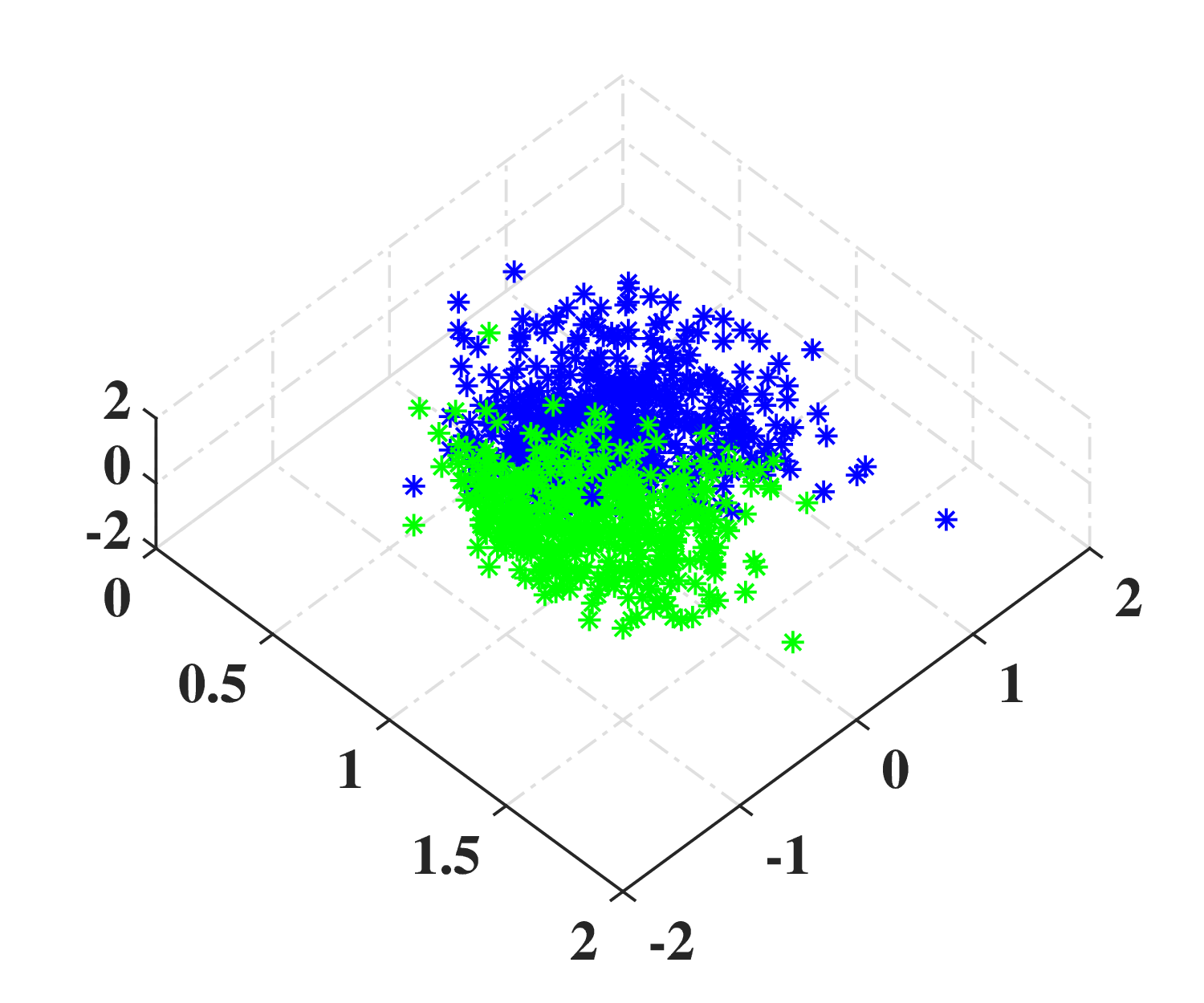}}
	\subfigure[]{\label{fig:Exp6f}\includegraphics[width = .3\textwidth]{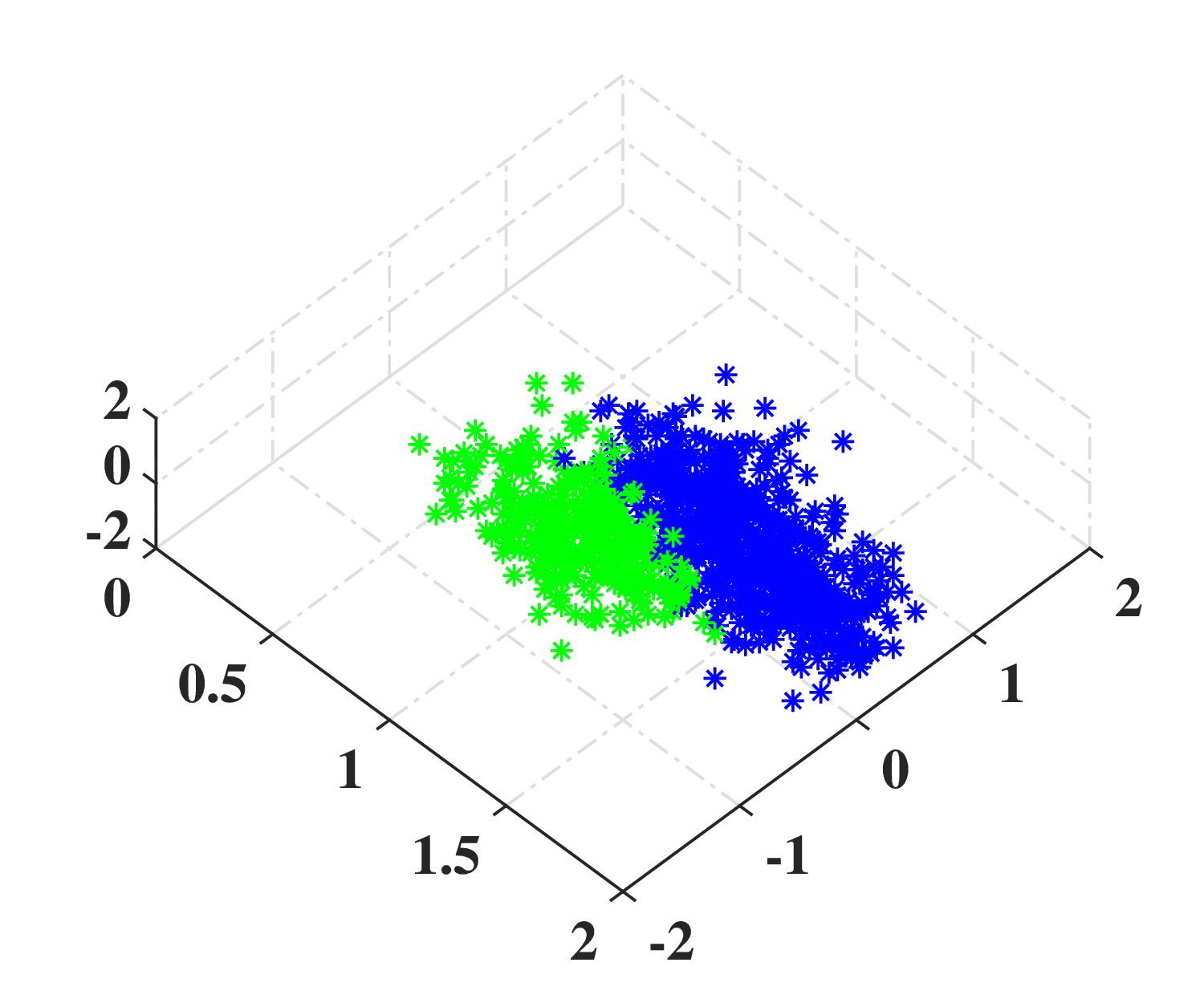}}
	\caption{Scatter plot of the first three features of (a) original spikes in Difficult1 dataset, (b) spikes reconstructed by WALM, (c) spikes reconstructed by SDNCS, (d) spikes reconstructed by AL1, (e) spikes reconstructed by BPDN, and (f) spikes reconstructed by BSBL, respectively. The features were extracted using wavelet decomposition. Points are colored according to the cluster to which they are assigned. The number of measurements is $M = 16$.}
	\label{fig:Exp6}
\end{figure*}

Fig. \ref{fig:Exp5} and Fig. \ref{fig:Exp6} show the three-dimensional (3D) projections of the first three features of reconstructed spikes using the five algorithms in the Easy1 and Difficult1 datasets, respectively. In all cases, the classification was done automatically with SPC and is represented in different colors. For the Easy1 dataset, we observe that it is possible to identify the three clusters clearly using the spikes reconstructed by all five algorithms. Furthermore, the features of spikes reconstructed by WALM and SDNCS are more accurate than that of the other algorithms, implying that WALM and SDNCS have better reconstruction performance. For the Difficult1 dataset, we observe that only the spikes reconstructed by WALM and SDNCS can be clearly classified, whereas the classification got failed using the spikes reconstructed by AL1, BPDN, and BSBL.

\begin{figure}[tbp]
	\centering
	\subfigure[]{\label{fig:Exp3a}\includegraphics[width = .5\textwidth]{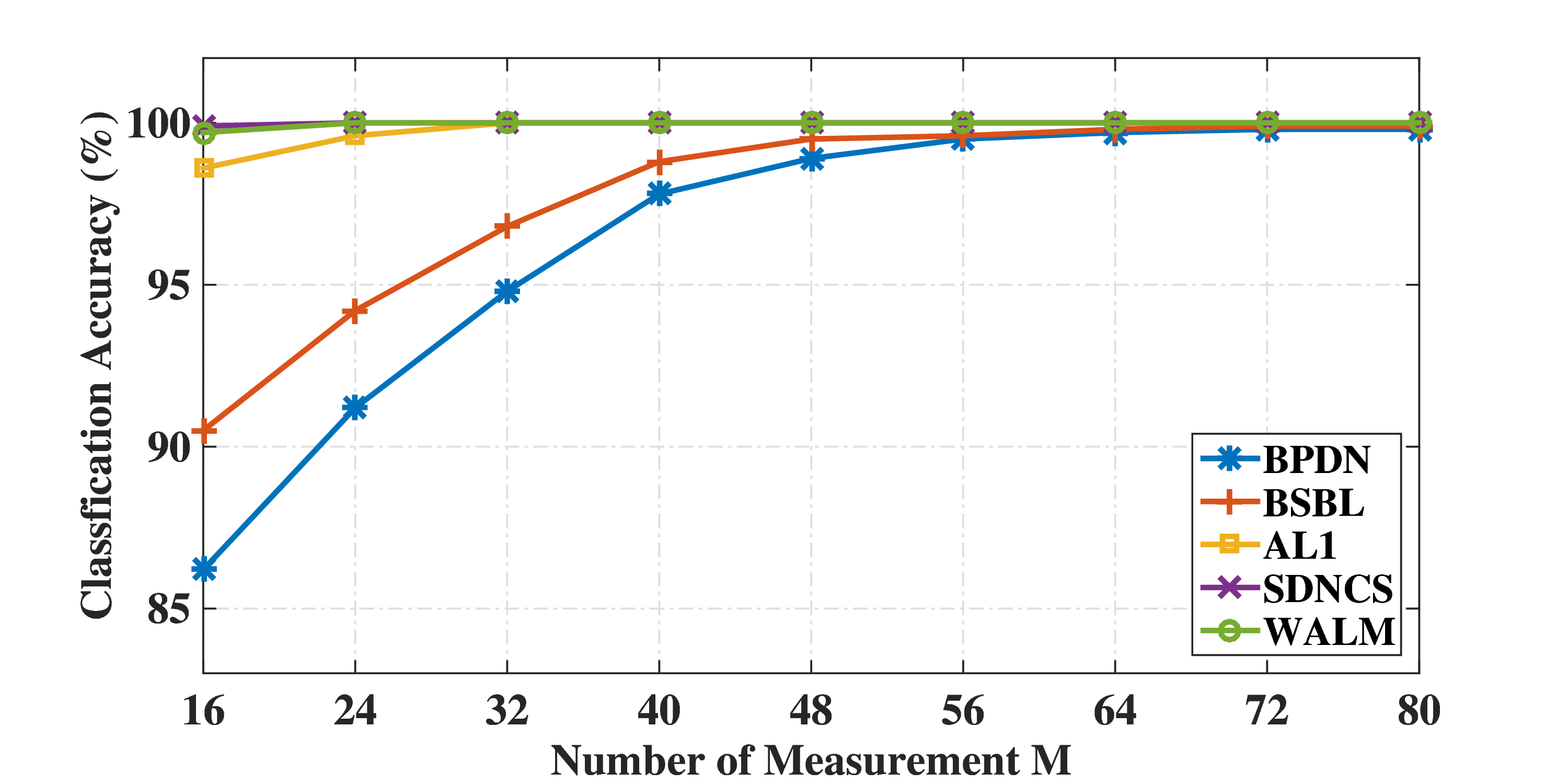}}
	\subfigure[]{\label{fig:Exp3b}\includegraphics[width = .5\textwidth]{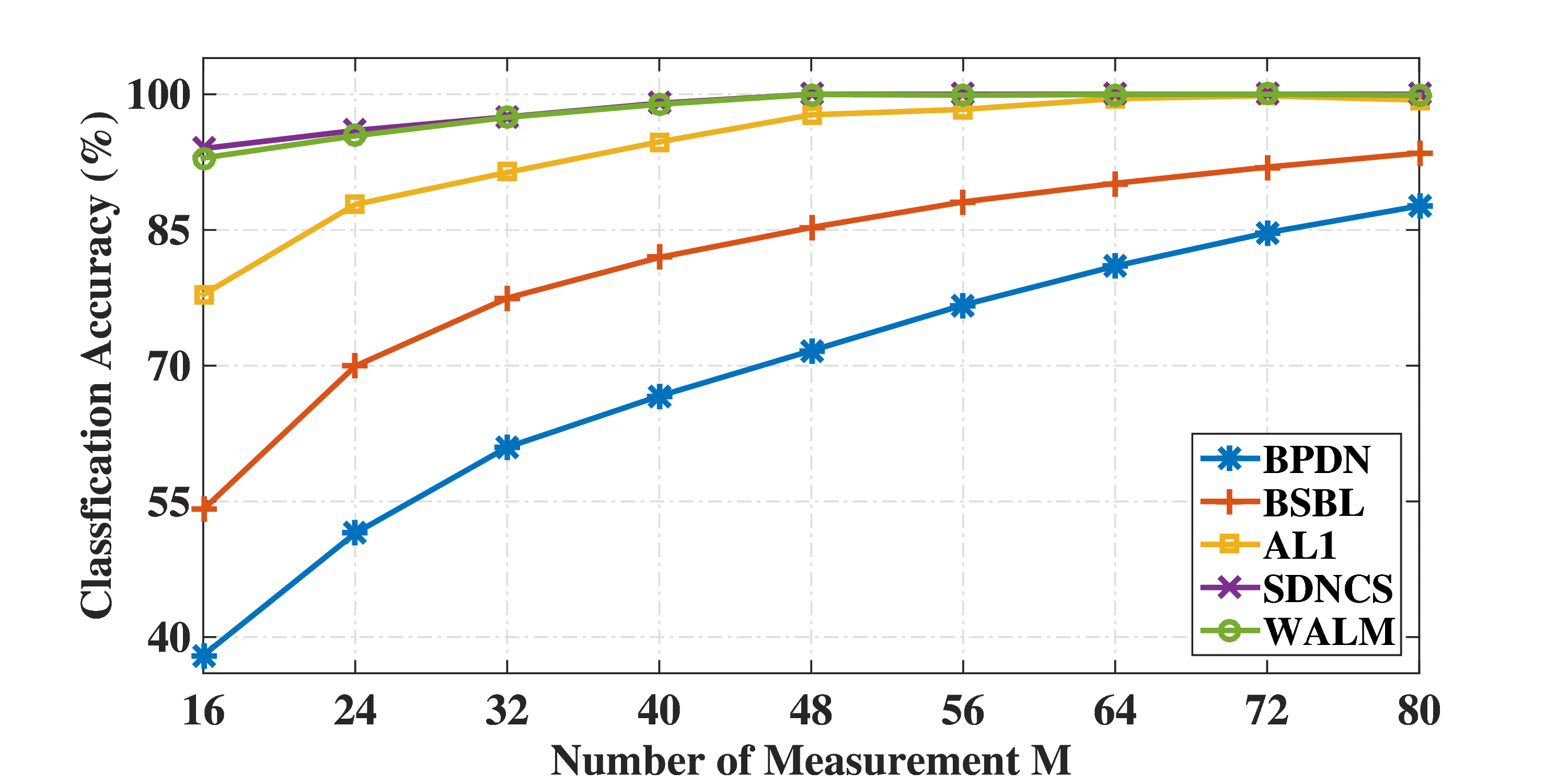}}
	\caption{Classification accuracy averaged over all spikes from (a) Easy1 dataset, (b) Difficult1 dataset, versus different numbers of measurements $M$ for BPDN, BSBL, AL1, SDNCS, and WALM, respectively.}
	\label{fig:Exp3}
\end{figure}

Spike classification accuracy was also used as a performance metric, calculated as a percentage of the total number of spikes correctly classified. The classification results were compared with the ground truth labels contained in the datasets. The spike classification results for Easy1 and Difficult1 are shown in Fig. \ref{fig:Exp3}. We observe that WALM and SDNCS outperform the other three algorithms for spike classification. Even the number of measurements is only 16, WALM can achieve above $99\%$ and above $92\%$ classification accuracy for Easy1 and Difficult1 datasets, respectively. Moreover, WALM provides a reliable solution which yields better reconstruction and classification performance with much fewer computational resource and pre-acquired data than SDNCS.

\section{Conclusion}
This paper proposed a novel compressed sensing method for implantable neural recording. The proposed method enforces sparsity of neural spikes not by the traditional synthesis model, but by the analysis model with a multiple fractional orders difference matrix as its analysis dictionary. Therefore, the pre-acquired data and computational resource for dictionary learning will be significantly reduced. Besides, by exploiting statistical priors of the analysis coefficients among difference orders, a weighted analysis $\ell_1$-minimization algorithm was proposed to reconstruct neural spikes. Experimental results proved the efficacy of the proposed method for neural signal reconstruction.

\bibliographystyle{IEEEtran}
\bibliography{bib_Implant}

\begin{thebibliography}{10}
\providecommand{\url}[1]{#1}
\csname url@samestyle\endcsname
\providecommand{\newblock}{\relax}
\providecommand{\bibinfo}[2]{#2}
\providecommand{\BIBentrySTDinterwordspacing}{\spaceskip=0pt\relax}
\providecommand{\BIBentryALTinterwordstretchfactor}{4}
\providecommand{\BIBentryALTinterwordspacing}{\spaceskip=\fontdimen2\font plus
\BIBentryALTinterwordstretchfactor\fontdimen3\font minus
  \fontdimen4\font\relax}
\providecommand{\BIBforeignlanguage}[2]{{%
\expandafter\ifx\csname l@#1\endcsname\relax
\typeout{** WARNING: IEEEtran.bst: No hyphenation pattern has been}%
\typeout{** loaded for the language `#1'. Using the pattern for}%
\typeout{** the default language instead.}%
\else
\language=\csname l@#1\endcsname
\fi
#2}}
\providecommand{\BIBdecl}{\relax}
\BIBdecl

\bibitem{stevenson2011advances}
I.~H. Stevenson and K.~P. Kording, ``How advances in neural recording affect
  data analysis,'' \emph{Nature neuroscience}, vol.~14, no.~2, pp. 139--142,
  2011.

\bibitem{berenyi2014large}
A.~Ber{\'e}nyi, Z.~Somogyvari, A.~J. Nagy, L.~Roux, J.~D. Long, S.~Fujisawa,
  E.~Stark, A.~Leonardo, T.~D. Harris, and G.~Buzs{\'a}ki, ``Large-scale,
  high-density (up to 512 channels) recording of local circuits in behaving
  animals,'' \emph{Journal of neurophysiology}, vol. 111, no.~5, pp.
  1132--1149, 2014.

\bibitem{schwarz2014chronic}
D.~A. Schwarz, M.~A. Lebedev, T.~L. Hanson, D.~F. Dimitrov, G.~Lehew, J.~Meloy,
  S.~Rajangam, V.~Subramanian, P.~J. Ifft, Z.~Li \emph{et~al.}, ``Chronic,
  wireless recordings of large-scale brain activity in freely moving rhesus
  monkeys,'' \emph{Nature methods}, vol.~11, no.~6, pp. 670--676, 2014.

\bibitem{yin2014wireless}
M.~Yin, D.~A. Borton, J.~Komar, N.~Agha, Y.~Lu, H.~Li, J.~Laurens, Y.~Lang,
  Q.~Li, C.~Bull \emph{et~al.}, ``Wireless neurosensor for full-spectrum
  electrophysiology recordings during free behavior,'' \emph{Neuron}, vol.~84,
  no.~6, pp. 1170--1182, 2014.

\bibitem{rodriguez2012low}
A.~Rodriguez-Perez, J.~Ruiz-Amaya, M.~Delgado-Restituto, and
  A.~Rodriguez-Vazquez, ``A low-power programmable neural spike detection
  channel with embedded calibration and data compression,'' \emph{Biomedical
  Circuits and Systems, IEEE Transactions on}, vol.~6, no.~2, pp. 87--100,
  2012.

\bibitem{chae2009128}
M.~S. Chae, Z.~Yang, M.~R. Yuce, L.~Hoang, and W.~Liu, ``A 128-channel 6 mw
  wireless neural recording ic with spike feature extraction and uwb
  transmitter,'' \emph{Neural Systems and Rehabilitation Engineering, IEEE
  Transactions on}, vol.~17, no.~4, pp. 312--321, 2009.

\bibitem{gosselin2009ultra}
B.~Gosselin and M.~Sawan, ``An ultra low-power cmos automatic action potential
  detector,'' \emph{Neural Systems and Rehabilitation Engineering, IEEE
  Transactions on}, vol.~17, no.~4, pp. 346--353, 2009.

\bibitem{gosselin2009mixed}
B.~Gosselin, A.~E. Ayoub, J.-F. Roy, M.~Sawan, F.~Lepore, A.~Chaudhuri, and
  D.~Guitton, ``A mixed-signal multichip neural recording interface with
  bandwidth reduction,'' \emph{Biomedical Circuits and Systems, IEEE
  Transactions on}, vol.~3, no.~3, pp. 129--141, 2009.

\bibitem{oweiss2007scalable}
K.~G. Oweiss, A.~Mason, Y.~Suhail, A.~M. Kamboh, and K.~E. Thomson, ``A
  scalable wavelet transform vlsi architecture for real-time signal processing
  in high-density intra-cortical implants,'' \emph{Circuits and Systems I:
  Regular Papers, IEEE Transactions on}, vol.~54, no.~6, pp. 1266--1278, 2007.

\bibitem{donoho2006compressed}
D.~L. Donoho, ``Compressed sensing,'' \emph{Information Theory, IEEE
  Transactions on}, vol.~52, no.~4, pp. 1289--1306, 2006.

\bibitem{candes2006near}
E.~J. Candes and T.~Tao, ``Near-optimal signal recovery from random
  projections: Universal encoding strategies?'' \emph{Information Theory, IEEE
  Transactions on}, vol.~52, no.~12, pp. 5406--5425, 2006.

\bibitem{zhang2014efficient}
J.~Zhang, Y.~Suo, S.~Mitra, S.~P. Chin, S.~Hsiao, R.~F. Yazicioglu, T.~D. Tran,
  and R.~Etienne-Cummings, ``An efficient and compact compressed sensing
  microsystem for implantable neural recordings,'' \emph{Biomedical Circuits
  and Systems, IEEE Transactions on}, vol.~8, no.~4, pp. 485--496, 2014.

\bibitem{suo2014energy}
Y.~Suo, J.~Zhang, T.~Xiong, P.~S. Chin, R.~Etienne-Cummings, and T.~D. Tran,
  ``Energy-efficient multi-mode compressed sensing system for implantable
  neural recordings,'' \emph{Biomedical Circuits and Systems, IEEE Transactions
  on}, vol.~8, no.~5, pp. 648--659, 2014.

\bibitem{zhang2013extension}
Z.~Zhang and B.~Rao, ``Extension of sbl algorithms for the recovery of block
  sparse signals with intra-block correlation,'' \emph{Signal Processing, IEEE
  Transactions on}, vol.~61, no.~8, pp. 2009--2015, 2013.

\bibitem{bulach2012evaluation}
C.~Bulach, U.~Bihr, and M.~Ortmanns, ``Evaluation study of compressed sensing
  for neural spike recordings,'' in \emph{Engineering in Medicine and Biology
  Society (EMBC), 2012 Annual International Conference of the IEEE}.\hskip 1em
  plus 0.5em minus 0.4em\relax IEEE, 2012, pp. 3507--3510.

\bibitem{bruckstein2009sparse}
A.~M. Bruckstein, D.~L. Donoho, and M.~Elad, ``From sparse solutions of systems
  of equations to sparse modeling of signals and images,'' \emph{SIAM review},
  vol.~51, no.~1, pp. 34--81, 2009.

\bibitem{becker2011nesta}
S.~Becker, J.~Bobin, and E.~J. Cand{\`e}s, ``Nesta: a fast and accurate
  first-order method for sparse recovery,'' \emph{SIAM Journal on Imaging
  Sciences}, vol.~4, no.~1, pp. 1--39, 2011.

\bibitem{candes2008restricted}
E.~J. Cand{\`e}s, ``The restricted isometry property and its implications for
  compressed sensing,'' \emph{Comptes Rendus Mathematique}, vol. 346, no.~9,
  pp. 589--592, 2008.

\bibitem{elad2007analysis}
M.~Elad, P.~Milanfar, and R.~Rubinstein, ``Analysis versus synthesis in signal
  priors,'' \emph{Inverse problems}, vol.~23, no.~3, p. 947, 2007.

\bibitem{nam2013cosparse}
S.~Nam, M.~E. Davies, M.~Elad, and R.~Gribonval, ``The cosparse analysis model
  and algorithms,'' \emph{Applied and Computational Harmonic Analysis},
  vol.~34, no.~1, pp. 30--56, 2013.

\bibitem{candes2011compressed}
E.~J. Candes, Y.~C. Eldar, D.~Needell, and P.~Randall, ``Compressed sensing
  with coherent and redundant dictionaries,'' \emph{Applied and Computational
  Harmonic Analysis}, vol.~31, no.~1, pp. 59--73, 2011.

\bibitem{peleg2013performance}
T.~Peleg and M.~Elad, ``Performance guarantees of the thresholding algorithm
  for the cosparse analysis model,'' \emph{Information Theory, IEEE
  Transactions on}, vol.~59, no.~3, pp. 1832--1845, 2013.

\bibitem{liu2015compressed}
Y.~Liu, M.~De~Vos, and S.~Van~Huffel, ``Compressed sensing of multichannel eeg
  signals: The simultaneous cosparsity and low-rank optimization,''
  \emph{Biomedical Engineering, IEEE Transactions on}, vol.~62, no.~8, pp.
  2055--2061, 2015.

\bibitem{chambolle2004algorithm}
A.~Chambolle, ``An algorithm for total variation minimization and
  applications,'' \emph{Journal of Mathematical imaging and vision}, vol.~20,
  no. 1-2, pp. 89--97, 2004.

\bibitem{keogh2001dimensionality}
E.~Keogh, K.~Chakrabarti, M.~Pazzani, and S.~Mehrotra, ``Dimensionality
  reduction for fast similarity search in large time series databases,''
  \emph{Knowledge and information Systems}, vol.~3, no.~3, pp. 263--286, 2001.

\bibitem{jiang2009new}
Y.~Jiang, T.~Lan, and D.~Zhang, ``A new representation and similarity measure
  of time series on data mining,'' in \emph{Computational Intelligence and
  Software Engineering, 2009. CiSE 2009. International Conference on}.\hskip
  1em plus 0.5em minus 0.4em\relax IEEE, 2009, pp. 1--5.

\bibitem{zhou2012new}
J.~Zhou, G.~Ye, and D.~Yu, ``A new method for piecewise linear representation
  of time series data,'' \emph{Physics Procedia}, vol.~25, pp. 1097--1103,
  2012.

\bibitem{yan2013approach}
C.~Yan, J.~Fang, L.~Wu, and S.~Ma, ``An approach of time series piecewise
  linear representation based on local maximum, minimum and extremum,''
  \emph{Journal of Information \& Computational Science}, vol.~10, no.~9, pp.
  2747--2756, 2013.

\bibitem{et1995some}
M.~Et and R.~{\c{C}}olak, ``On some generalized difference sequence spaces,''
  \emph{Soochow Journal of Mathematics}, vol.~21, no.~4, pp. 377--386, 1995.

\bibitem{quiroga2004unsupervised}
R.~Q. Quiroga, Z.~Nadasdy, and Y.~Ben-Shaul, ``Unsupervised spike detection and
  sorting with wavelets and superparamagnetic clustering,'' \emph{Neural
  computation}, vol.~16, no.~8, pp. 1661--1687, 2004.

\bibitem{chen2012design}
F.~Chen, A.~P. Chandrakasan, and V.~M. Stojanovi{\'c}, ``Design and analysis of
  a hardware-efficient compressed sensing architecture for data compression in
  wireless sensors,'' \emph{Solid-State Circuits, IEEE Journal of}, vol.~47,
  no.~3, pp. 744--756, 2012.

\bibitem{zigel2000weighted}
Y.~Zigel, A.~Cohen, and A.~Katz, ``The weighted diagnostic distortion (wdd)
  measure for ecg signal compression,'' \emph{Biomedical Engineering, IEEE
  Transactions on}, vol.~47, no.~11, pp. 1422--1430, 2000.

\bibitem{donoho2008sparselab}
D.~Donoho, V.~Stodden, and Y.~Tsaig, ``Sparselab. 2007,'' \emph{See
  http://sparselab. stanford. edu..(Accessed January 23, 2014.)}, 2008.

\bibitem{grant2008cvx}
M.~Grant, S.~Boyd, and Y.~Ye, ``Cvx: Matlab software for disciplined convex
  programming,'' 2008.

\bibitem{jolliffe2002principal}
I.~Jolliffe, \emph{Principal component analysis}.\hskip 1em plus 0.5em minus
  0.4em\relax Wiley Online Library, 2002.

\end{thebibliography}

\end{document}